\documentclass{article}
\usepackage{amssymb}
\usepackage{amsmath}
\usepackage[T1]{fontenc}
\usepackage[latin1]{inputenc}
\usepackage{graphicx}
\usepackage{revsymb}

\newcommand{\dsum} {\displaystyle\sum}

\def\mnras{{MNRAS}}
\def\pra{{Phys.~Rev.~A}}

\def\nat{{Nature}}

\def\physrep{{Phys.~Rep.}}

\input{tcilatex}
\begin{document}

\title{Dipole blockade in a cold Rydberg atomic sample}
\author{Daniel Comparat and Pierre Pillet}
\date{\today}
\maketitle

\begin{abstract}
We review here the studies performed about interactions in an assembly of
cold Rydberg atoms. We focus more specially the review on the dipole-dipole
interactions and on the effect of the dipole blockade in the laser Rydberg
excitation, which offers attractive possibilities for quantum engineering.
We present first the various interactions between Rydberg atoms.\ The laser
Rydberg excitation of such an assembly is then described with the
introduction of the dipole blockade phenomenon. We report recent experiments
performed in this subject by starting with the case of a pair of atoms
allowing the entanglement of the wave-functions of the atoms and opening a
fascinating way for the realization of quantum bits and quantum gates. We
consider then several works on the blockade effect in a large assembly of
atoms for three different configurations: blockade through electric-field
induced dipole, through Förster resonance and in van der Waals interaction.
The properties of coherence and cooperativity are analyzed. Finally, we
treat the role of dipole-dipole interactions between Rydberg atoms
responsible for Penning ionization.\ The perturbation of the dipole blockade
by ions and the evolution of the Rydberg towards an ultracold plasma are
discussed.
\end{abstract}

\affiliation{Laboratoire Aimé Cotton, CNRS, Univ Paris-Sud, Bât. 505, 91405
Orsay, France}



\section{Introduction to Rydberg atoms}

The field of cold Rydberg atoms or molecules knows today an intense
international activity, motivated by both fundamental scientific research and
applications. The main attractivity of Rydberg systems is linked to the
existence of huge electric dipolar momenta, offering various possibilities
such as manipulation of the Rydberg particles with small gradients
of electric field \cite{2008PhRvL.100d3001H}, or control of interactions between two particles at a macroscopic
distance \cite{review_gall_pillet}.  By using highly excited Rydberg gases with strong and long-range
inter-particle interactions, many complex quantum systems can be achieved
with a full control of the dynamics, at the crossing of different areas as
solid state or plasma physics. The effects of the long-range dipole-dipole interactions between Rydberg atoms   have been recently reviewed in 
\cite{review_gall_pillet}.
 The Rydberg excitation offers efficient tools
for a quantum engineering especially for the entanglement of neutral particles (see the review in reference \cite{2009arXiv0909.4777S}).\ A
demonstrated example is the dipole blockade \cite{2001PhRvL..87c7901L} of the laser excitation, effect
on which is focused this review and which is the result of the long-range
interatomic dipole-dipole interactions shifting the Rydberg energy levels and preventing their laser excitation. 

In the eighties, the atomic highly excited or Rydberg states (with $n$, $%
\ell $, $m_{\ell }$ respectively principal, orbital angular and magnetic
quantum numbers) were already extensively studied for their exaggerated
properties, as for instance their giant size ($\sim 1\mu $m for $n\sim 100$)
ranging as $n^{2}$ in atomic units (Bohr radius, $a_{0}$) \cite%
{gallagher1994}. Such states are so sensitive to weak perturbations
that they offer the opportunity to explore phenomena inaccessible for other
physical systems in laboratory experiments. Rydberg atoms have huge
polarizabilities leading to extreme collisional properties of room
temperature atoms, in particular, large cross sections and long interaction
times. 

The effect of the dipole-dipole interaction displays very clearly in
the interactions between Rydberg atoms having high principal quantum number $%
n$, whose dipole moments scale as $n^{2}$. Due to the strong and long-range
interactions, a Rydberg gas offers the opportunity for studying $N$%
-body, many-body cooperative or collective effects \cite%
{2006PhRvA..73c2725C,2008PhRvL.100d3002T,2008PhRvL.100l3007R,review_gall_pillet}.
A key point is that it is experimentally possible to tune, for instance by  micro-wave or external electric fields, the Rydberg energy levels to achieve a resonant condition. One of the most interesting possibility, on which 
this review is focused on,
 is when 
the two Rydberg atoms  exchange resonantly internal energy by making
transitions between Rydberg levels in a $r_1+r_2\rightarrow r'_1+r'_2$ iso-energetic reaction. This configuration of the collisional
process, a so called Förster
resonance \cite{2005JPhB...38S.309W}, has been first demonstrated by Thomas F.\ Gallagher's group  (Virginia University, USA) \cite{1981PhRvL..47..405S,1992PhR...210..319G} in a thermal ensemble and later together with  Pierre Pillet's group (Orsay, France) in a cold atomic gas \cite{anderson1998,mourachko1998}.

We concentrate this review to the 
study of cold Rybderg gases. Such study started in Pierre Pillet
and Thomas Gallagher's groups  in 1998 with
already signature of $N$-body effects in order to interpret the system as a
frozen gas with properties quite similar to those of an artificial solid 
\cite{mourachko1998,anderson1998}.  The first
modeling of cold dipolar interaction was thus based on the so called "frozen
Rydberg gas" hypothesis \cite{1999PhyD..131..125A}. However, it has very
soon be recognized that a dipole gas does not stay a long time as a frozen
gas. The whole dynamics of the system necessitates to take into account the
dipole forces between close Rydberg atoms \cite{fioretti1999a}.
These forces are at the origin of the creation of ions by Penning
ionization.\ An extreme and impressive effect of this dynamics is the
ultra-cold plasma formation \cite{vitrant1982,kulin1999,ref2000PhRvL854466R}.

The dipole-dipole interaction between Rydberg atoms creates a shift
depending on the interatomic distance, giving the asymptotic dependence of
the molecular potential curves at large distance from which the forces derive. This energy shift can also
allows to control the excitation of the neighboring atoms around a previously
excited one, as proposed in 2000-2001 \cite%
{2000PhRvL..85.2208J,2001PhRvL..87c7901L}. The shift created by the
dipole-dipole interactions allow for instance a single Rydberg excitation in
a large atomic sample, creating a Fock state. The dipole blockade 
regime for two atoms have been recently observed in the collective single
excitation of a pair of individually trapped atoms \cite%
{2009NatPh...5..115G,2010PhRvL.104a0503I} .\ Such a result opens the way to
the control of few-atom sample. Several previous studies in many-atom
sample have displayed a large variety of phenomena:
absorption, emission or ionization by black-body radiation, cooperative
spontaneous emission with limited superradiance, collective excitation, many
body interactions, dynamical dipolar forces, excitation transfer, dipole
orientation effects, fine or hyperfine atomic effects, random position of
atoms or ions effects in ultracold plasmas... Finally, with the great
experimental control obtained, a many-atomic sample can probably be employed
as quantum simulator to answer some challenging open questions of condensed
matter \cite{2007AdPhy..56..243L,2009Sci...326..108B}.

The paper is organized as following. In the section \ref{prop_Ryd}, we
recall a few of the main characteristics of Rydberg atoms. In section \ref%
{int_Ryd} we discuss the Rydberg-Rydberg interactions and more precisely the
dipole-dipole ones. The case of the retardation effects  is described
carefully.   In the section \ref{sec_dip_two_at}, we  expose a first
consequence of the dipole-dipole interaction, the dipole blockade of the
Rydberg laser excitation of cold atoms. We also discuss recent experiments using
simply two atoms, and succeeding to entangle them through a collective
excitation in the blockade regime. In section \ref{sec_dip_many_at} we
detail the condition for the observation of the blockade in the case of a
large sample in the configurations of the van der Waals interaction, of a
permanent dipole induced by a static electric field, and of a Förster
resonance. In the section \ref{sec_dyn}, we discuss the limit of the picture
of the frozen Rydberg gas, through the presence of the forces exerted
between two close Rydberg atoms in dipole-dipole interaction. We report the
possibility to control the attractive or repulsive character of these
forces, for preventing the Penning ionization particularly important in the
evolution of cold Rydberg gases toward ultracold plasmas. The section \ref%
{sec_con} is the conclusion.

\section{Properties of Rydberg atoms}

\label{prop_Ryd}

The properties of Rydberg atoms can be considered as really exaggerated
compared to "classical" atoms \cite{gallagher1994}. We have already
mentionned the size, we underline the following characteristics

\begin{enumerate}
\item The binding energy is very small.

\item The radiative lifetime is very long.

\item The dipole matrix elements are big.

\item They are very sensitive to electric fields.
\end{enumerate}

First, the binding energy ($\sim 300\,$GHz for $n\sim 100$) is very small $E=%
\frac{e^{2}}{a_{0}}\frac{m-m_{e}}{m}\frac{1}{2(n-\delta _{\ell })^{2}}$,
where $e^{2}=\frac{q_{e}^{2}}{4\pi \varepsilon _{0}}$, $m_{e}$ is the
electron mass, $m$ the mass of the atom, and $q_{e}$ the electron charge. $n^{\ast }=n-\delta _{\ell }$ is the effective quantum number where $%
\delta _{\ell }$ is called the quantum defect and depends mainly only of the
orbital angular momentum $\ell $ of the valence electron. The low $\ell $
angular states orbits penetrate the core and are therefore the most
perturbated ones with large quantum defect. For instance the $\ell =0$ ($s$
states) have a quantum defect $\delta _{0} = 4.052$ in the cesium case.
Spin-orbit effect are still present, even at high $n$ values, e.g. a fine
structure splitting on the order of $100\,$MHz exists for Rb($p$) levels at $%
n=100$ \cite{2008PhRvA..77c2723W}.

Second, the radiative lifetime is very long, on the order of several
microseconds. A useful very simplified formula to estimate the lifetime is \cite%
{2005JPhB...38.1765H} $n^3 (\ell + 1/2)^2 10^{-10}\,$s. Note that the lifetime rapidly increases as $n$ or $\ell$ increases. This formula does not include
the blackbody effect \cite{2009PhRvA..79e2504B} which, for a temperature $T$, increase the decay rate by roughly  $2 \times 10^7 \frac{T(K)}{300K}
n^{-2}\,(s^{-1})$ \cite{1981PhRvA..23.2397F}.

Third, as mentioned in the introduction, the radial dipole matrix element $%
R_{n\ell }^{n^{\prime }\ell \pm 1}=\langle n\ell |r|n^{\prime }\ell \pm
1\rangle $ are big\footnote{%
This has not to be confused with the reduced matrix element $\langle n\ell
||r||n^{\prime }\ell \pm 1\rangle =\sqrt{2\ell ^{\prime }+1}C_{\ell
+10,10}^{\ell 0}\langle n\ell |r|n^{\prime }\ell \pm 1\rangle $, $%
C_{j_{1}m_{1},j_{2}m_{2}}^{jm}$ being the standard Clebch-Gordan coefficient. For instance, with $\ell ^{\prime
} = \ell \pm 1$, $\left\langle n\ell m\left\vert q_{e}z\right\vert n^{\prime }\ell ^{\prime
}m \right\rangle =\sqrt{\frac{(\ell +\ell ^{\prime }+1-2m)(\ell
+\ell ^{\prime }+1+2m)}{4(2\ell +1)(2\ell ^{\prime }+1)}}R_{n\ell
}^{n^{\prime }\ell ^{\prime }}$,  where $z$ is
the coordinate along the quantization
axis.}, on the order of $a_{0}n^{2}$. A useful semi classical expression \cite%
{1995JPhB...28.4963K,1988PhRvA..37.1885O} is: 
\begin{equation*}
R_{n\ell }^{n^{\prime }\ell \pm 1}\approx \frac{a_{0}}{s}\frac{n_{c}^{5}}{%
(n^{\ast }n^{\prime \ast })^{3/2}}\left[ J_{-s}^{\prime }(es)\pm \sqrt{%
e^{-2}-1}\left( J_{-s}(es)-\frac{\sin (\pi s)}{\pi s}\right) +(1-e)\frac{%
\sin (\pi s)}{\pi }\right] 
\end{equation*}%
where  $%
s=n^{\prime \ast }-n^{\ast }$, $n_{c}^{3}=\frac{2(n^{\ast }n^{\prime \ast
})^{2}}{n^{\ast }+n^{\prime \ast }}$, $e^{2}=1-\left( \frac{\ell +\ell
^{\prime }+1}{2n_{c}}\right) ^{2}$ and $J_{s}(z)$ is the Anger function ($%
J_{s}^{\prime }$ its derivative). This expression is valid to high accuracy
(below $5$ percent)  for low angular momentum ($\ell <9$) states 
\cite{1998JPhB...31..963L}. It is also correct when 
$n^{\prime \ast }=n^{\ast }$ ($s\rightarrow 0$) and extension to the continuum states is
straightforward, the only change is on the normalization factor \cite%
{1994JPhB...27..461D,hoogenraad1998}. The main message is that the electric
dipole transition matrix can acquire huge values, $\sim n^{2}q_{e}a_{0}$,
where the atomic unit of electric dipole moment is $-q_{e}a_{0}=8.48\times
10^{-30}\,$C.m$=2.54\,$Debye.

\begin{figure}[h!]
\centering
\resizebox{0.9\textwidth}{!}{
		\rotatebox{-90}{
		\includegraphics*[43mm,72mm][152mm,226mm]{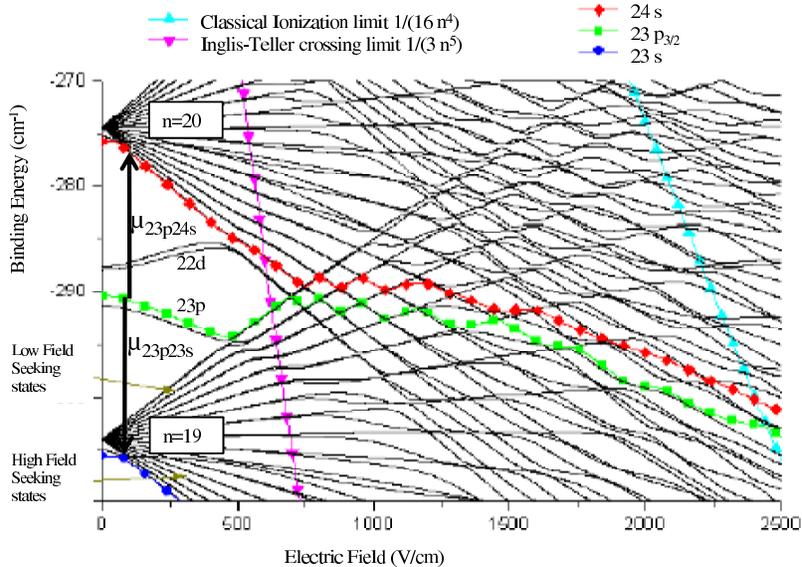}
		}}
\caption{Energy levels of the cesium Rydberg state $n \ell j m_j$ with $%
|m_j|=1/2$ versus the electric field. The fine structure of the $23 p$ level
is visible. The Förster resonance $23p +23p \rightarrow 23s+24s$ is resonant for a field of $\sim 80\,$V/cm.}
\label{fig:Stark}
\end{figure}

Forth, the Rydberg states are very sensitive to electric fields. The system
of the hydrogen atom in the presence of an electric field is well known to
have analytical solution.\ The effect of the electric field is to lift the $%
\ell $-degeneracy of the $n$\ manifold.  In the case of alkali atoms, the
behavior is not different for high angular momentum states, while the low
ones with large quantum defect have a quadratic energy behavior versus the
electric field.\ The standard way to calculate the external electric field effect, i.e. the
effect of the $-\vec{\mu}.\vec{F}$ Stark Hamiltonian, where $\vec{\mu}=q_{e}%
\vec{r}$ is the dipole operator of one atom, is to follow the work in
reference \cite{PhysRevA.20.2251}. A typical result is shown in figure \ref%
{fig:Stark}. If a (external or due to ions) static electric field $\vec{F}$
is present the Rydberg states are mixed, creating a permanent dipole moment
for the Rydberg atoms.
Adding a small electric field is therefore a very
simple way to control the Rydberg-Rydberg interaction.

For small dc electric fields, i.e. when the dipole coupling is of the same
order as the energy difference  of the nearest state with allowed
dipolar transition, the Stark effect is quadratic and the shift can be
calculated with a reasonable approximation in a two level model. As
illustrative example, in the cesium, this case concern the $r = n\ell
=np$ state is mainly mixed with the $r^{\prime } = n^{\prime }\ell ^{\prime }=(n-1)d$
state (the closest energetically nearby state with allowed dipole transition
see figure \ref{fig:Stark}). 
 For simplicity of the writing we do not express here any
dependence of $j$ the total angular momentum, neither of $m_j$ (or $m_\ell$)
its projection along a chosen axis the external field or the internuclear
axis. Depending on the experimental conditions (laser polarization mainly),
the projection of the angular momentum can be controlled or not.

We denote by $\vec \mu _{r r^{\prime
}}= \langle r |\vec \mu |r'\rangle$ the transition dipole moment of the $|r\rangle \rightarrow |r'\rangle$ transition. So the new eigenstates ($|r (\vec{F})\rangle $,$|r^{\prime }(\vec{F})\rangle $) are given by the diagonalization of the
Hamiltonian matrix: 
\begin{equation}
\left( 
\begin{array}{ccc}
E_r & -\mu _{r r^{\prime }}F &  \\ 
-\mu _{r r^{\prime }}F & E_{r^{\prime }}  & 
\end{array}%
\right)  \label{eq_champ}
\end{equation}%
where $E_{r}$ and $E_{r^{\prime }} = E_r + \hbar \omega_{r r'}$ are the energies of states $%
r $ and $r^{\prime }$ in absence of an electric field. We introduce the scale parameter $\vartheta $ characterizing the dipole
coupling for each level $r $ by $\tan \vartheta =\frac{\mu_{r r^{\prime }}F}{\hbar \omega_{r r'} /2}$. 
The atom in state $|r (\vec{F})\rangle $ acquires a classical
permanent electric dipole $\vec{\mu}_r (\vec{F})$ aligned along the local
electric field $\vec{F}$ with value $\mu _{r }(F)=\langle r (\vec{F}%
)|q_{e}z| r (\vec{F})\rangle =\mu _{r r^{\prime }}\sin \vartheta $%
. The resulting shift in energy for $r $ is $ \hbar \Delta _{r }(\vec{F})=%
\frac{\hbar \omega_{r r'}}{2}(1-\sqrt{1+\tan ^{2}(\vartheta )})$.

In the case of non hydrogen atoms as alkali ones, another important feature
of the effect of the electric field different from the hydrogen is the
avoided crossings between the Stark levels of two different manifold.\ As a
consequence the electric-field ionization of the Rydberg atoms depends on
the risetime of the pulsed ionization field and correspond to the ionization
classical limit. Consequently, an electric field of $3.21 \times \left(\frac{100}{n} \right)^4 \,$%
V/cm is usually enough to ionize a Rydberg state. 

\section{Interaction between two close Rydberg atoms}

\label{int_Ryd}

Rydberg studies are often done in presence of an external electric field,
either to ionize the atoms, to induce a permanent dipole moment or to study
what is called a (resonant) excitation transfer.

Detail study of van der Waals and dipole-dipole energy shifts of pairs of
interacting Rydberg atoms for different quantum numbers $n , l , j ,$ and $%
m_j$, taking into account a large number of perturbing states has been
performed, and useful formula have been given in references \cite%
{2005JPhB...38S.295S,2007PhRvA..75c2712R,2007PhRvA..75c9902R,2008PhRvA..77c2723W}%
. We would like here simply to introduce some of the key notions.

The electrostatic interaction hamiltonian between two atoms, the first one
(core $A$ electron $1$) and the second (core $B$ electron $2$) separated
by $\vec{R_{ A B}}=R\vec{n}$ is, with obvious notations: $H = {\frac{ e^2 }{R }} - {\frac{ e^2 }{%
r_{A 2} }} - {\frac{ e^2 }{r_{B 1} }} + {\frac{ e^2 }{r_{12} }} $. Using $%
(\parallel \vec R + \vec r \parallel^2 )^{-1/2} = {\frac{1 }{R}} \left( 1 - {%
\frac{\vec n.\vec r }{R}} + {\frac{3 (\vec n.\vec r)^2 - r^2 }{2 R^2}} +
\cdots \right) $, the leading term in the asymptotic expansion (i.e. large $%
R $), is the so called dipole-dipole interaction and is found to be: 
\begin{equation}
H_{1 2} = {\frac{e^2 }{R^3}} \left( \vec r_{A 1}.\vec r_{B 2} - 3 (\vec r_{A 1}
. \vec n ) (\vec r_{B 2} . \vec n ) \right) = {\frac{ \vec \mu_1.\vec
\mu_{2} - 3 (\vec \mu_{1} . \vec n ) (\vec \mu_{2} . \vec n )  }{4\pi
\varepsilon_0 R^3}} .  \label{int_dipol_dip} 
\end{equation}
Due to the long range interaction electron $1$ is always close to core $A$ and electron $2$ to core $B$ so there is not exchange interaction, i.e. no spatial overlap between the wave function of electron $1$ and $2$. We  thus think that there is no  confusion in the notation $\vec \mu_{1}$ which is the dipole operator for the atom $A$.

Before using this hamiltonian to calculate potential energy curves between Rydberg atoms, we would like
to study this hamiltonian in a (semi-)classical point of view which will allow us to use
 the retardated potentials.

In the appendix (section \ref{appendix}), see also
\cite{2005quant.ph..9184Y,2008PhRvA..77c3844B},
 we shall show the equivalence of this
semi-classical approach, with the
establishment of the master equation of the evolution of a pair of two-level
atoms coupling with the electromagnetic field of the vacuum.\ The master
equation is obtained in the Born-Markov approximation.\ It takes into
account several effects blind to the semi-classical picture namely: the cooperative spontaneous emission or superradiance demonstrating the strong link between the two
non-separable effects.

\subsection{Semi-classical derivation of the dipole-dipole hamiltonian interaction}

\subsubsection{Instantaneous dipole-dipole interaction}

Classically speaking,
a particle, such as one atom in the $|r(\vec F)\rangle$ state, which possesses a permanent electric dipole moment, $%
\vec{\mu }$, generates an electric field.\ At a (vectorial) distance, $R%
\vec{n}$, the expression of the field is $
\vec{E}=\frac{1}{4\pi \varepsilon _{0}}\left[ 3%
\left( \vec{n}.\vec{\mu }\right) \vec{n} -\vec{\mu }%
\right] \frac{1}{R^{3}}.
$
For two particles, in states  $|r_1(\vec F)\rangle$ and $|r_2(\vec F)\rangle$, separated by the distance, $R\vec{n}$, the
energy of the system is given by
$
V_{12}=-\vec{E}_{1}.\vec{\mu }_{2}=-\vec{\mu }_{1}.\vec{E}_{2}=\frac{ 
\vec{\mu }_{1}.\vec{\mu }_{2}- 3\left( \vec{%
\mu }_{1}. \vec{n} \right) \left( \vec{\mu }_{2}. 
\vec{n} \right) }{4\pi \varepsilon _{0} R^{3}}.
$
This results is exactly the quantum mechanical one given by Eq. (\ref{int_dipol_dip}), i.e.
$V_{12} = \langle r_1(\vec F),r_2(\vec F) | H_{12}|r_1(\vec F),r_2(\vec F)\rangle$. More generally
the hamiltonian $ H_{12}$  is simply $ V_{12}$
 with the vectors dipoles $\vec{\mu }_{1},\vec{\mu }_{2}$ becoming operators.

In the case where the imposed external field $\vec F$ is stronger than all other fields, such as the sum of the  fields created by the dipoles, all dipoles are aligned along the same field
$\vec F$.
 If we
define the angle $\theta $ as $\cos \theta =( \vec{n}.%
\vec{F })/F $ we thus have%
\begin{eqnarray}
V_{12}=\frac{\mu_1 \mu_2}{4 \pi \epsilon_{0} R^{3}} \left( 1-3\cos ^{2}\theta \right)  \label{eq:dipdip2}
\end{eqnarray}

This $1-3\cos ^{2}\theta$ angular dependence has
been observed experimentally by the Noel's group in \cite%
{2004PhRvL..93o3001C} using a quasi one dimensional cylindrical atomic
sample which could be rotated with respect to the external electric field.

\subsubsection{Retardated dipole-dipole interaction}

In absence of electric field, the Rydberg atoms have no permanent electric
dipole. However they can still interact through exchange of excitation corresponding to the
existence of a transition dipole from state $|n \ell \rangle = |r \rangle$ to state $ | n' \ell' \rangle = |r' \rangle$; the typical example being the  $n s \rightarrow n p$ transition of an alkali atom.

Classically speaking, the transition $r  \rightarrow r'$, with dipolar matrix element $\vec \mu$ and transition energy $\hbar \omega = \hbar \omega_{r r'} = E_{r'} - E_{r}$
corresponds,
using the complex representation,
to an oscillating electric dipole
momentum, $\vec{\mu} \exp \left( -i \omega t\right) $. This dipole 
radiates an electromagnetic field, at a vectorial distance, $R\vec{n}$, 
$\vec{E} \left( t\right) = \vec{E} ( \vec{R}) \exp \left( -i \omega t\right) $, with 
\begin{equation}
\vec{E} ( \vec{R}) =\frac{\exp \left(
ikR\right) }{4\pi \varepsilon _{0}}\left\{ \left[ k^{2}\left( 
\vec{n}\times \vec{\mu }\right) \times 
\vec{n}\right] \frac{1}{R}+\left[ 3\left( 
\vec{n}.\vec{\mu }\right) \vec{n} -\vec{\mu }%
\right] \left( \frac{1}{R^{3}}-\frac{ik}{R^{2}}\right) \right\} 
\label{eq_un_sur_R}
\end{equation}
where $k=\omega/c$ is the wave number
of the light.
Thus in the reaction $r_1 + r_2 \rightarrow r'_1 + r'_2$, the electromagnetic field 
$\vec{E}_{1}( t) =\vec{E}%
_{1}( \vec{R}) \exp \left( -i\omega_{r_1 r'_1} t\right) $,
radiated  by the atom $%
1$ interacts with the dipole $\vec{\mu }_{2}\left( t\right) = \vec{\mu}_2 \exp \left( -i \omega_{r_2 r'_2} t\right) $ to give the interaction energy $
-\vec{E}_{1}( t) .%
\vec{\mu }_{2}( t) = - \vec{E}%
_{1}( \vec{R}) . \vec{\mu}_2 \exp \left( -i (\omega_{r_1 r'_1} + \omega_{r_2 r'_2}) t\right) $.
The phase indicates the energy mismatch for the transition and $- \vec{E}%
_{1}\left( R\vec{n}\right) . \vec{\mu}_2 $ is the energy in complex notation, so 
the (real)  energy is given by
\begin{eqnarray}
H_{12} & = &\func{Re}\left[ -\vec{E}_{1} (\vec R) .%
\vec{\mu }_{2} \right]=\frac{1}{4\pi \varepsilon _{0}}\Big\{ 
 \left[ \vec{\mu }_{1}.%
\vec{\mu }_{2} - 3\left( \vec{n}.\vec{\mu }%
_{1}\right) \left( \vec{n}.\vec{\mu }%
_{2}\right) \right] \left( \frac{\cos (k R)}{R^{3}}+\frac{k \sin (k R)%
}{R^{2}}\right) + \label{H_12_ret}
 \\
&& \left[ \vec{\mu }_{1}.%
\vec{\mu }_{2} - \left( \vec{n}.\vec{\mu }%
_{1}\right) \left( \vec{n}.\vec{\mu }%
_{2}\right) \right]  \frac{k^{2} \cos (k R)}{R}
\Big\}  \nonumber
\end{eqnarray}
where $k=k_1=\omega_{r_1 r'_1}/c$ is the wave number
of the light. 
We could have also calculate the energy by using  $\func{Re}\left[ -\vec{\mu }_{1} . \vec{E}_{2} 
 \right]$, which differs at first glance  because $k$ would have been 
 $k_2=\omega_{r_2 r'_2}/c$. But, due to energy conservation $ \omega_{r_1 r'_1} + \omega_{r_2 r'_2} \sim 0$, only near resonant term would contribute in the final evaluation so  $k_1\sim -k_2$. We note  $\lambda = 2\pi/|k|$
the  wavelength of the transition.

For a
short distance, $R<\lambdabar =\lambda /2\pi $, the hamiltonian
corresponds essentially to the electrostatic dipole-dipole interaction calculated previously in Eq. (\ref{int_dipol_dip}), and which can thus be interpreted as an exchange of
excitation between the two atoms (as an exchange of a virtual photon).

For a large distance, $R>\lambda $, only the term
in $1/R$ of the radiated electromagnetic field is important.

  It is interesting to consider the order of magnitude
of the different parameters in the case of the electric-dipolar transition $%
np\longrightarrow ns$ of an alkali atom. The difference between the quantum
defects of the $s$ and $p$ states for any alkali atom (except lithium) is $%
0.5$. The wave-length associated to the transition is therefore $\lambda
\sim 9$ cm (resp. 2, 0.6, 0.25, 0.07 ) and $\lambdabar =$ $\lambda /2\pi
\sim 1.5$ cm (resp. 0.3, 0.09,0.04, 0.01) for $n\sim 100$ (resp.\ 60, 40,\
30,\ 20). These values should to be compared to the size of the cold atomic
sample usually of the order of 0.1\ cm, which exceeds $\lambda $ \ for $n>22$
and $\lambdabar $ \ for $n>40$. At a distance of 1$\mu $m, the dipole-dipole
coupling is of the order of 80\ GHz  (resp.\ 10, 2,\ 0.65,\ 0.13) for $n\sim
100$ (resp.\ 60, 40,\ 30,\ 20).

Finally, to have the quantum mechanical equivalent of Eq. (\ref{H_12_ret}) we have to proceed as previously, i.e. to replace the dipole vectors by operators. For the atom $i$, 
$\vec{\mu }_i^{r_i  r'_i}= \left\langle r_i \right\vert \vec{d}\left\vert r'_i \right\rangle  \left\vert r'_i \right\rangle \left\langle
r_i \right\vert$ is
the dipolar operator associated to the transition $r_i \rightarrow r'_i$  of energy $\hbar \omega_{r_i r'_i} = E_{r'_i} - E_{r_i}$. 
If
we consider the reaction $r_1 + r_2 \leftrightarrow r'_1 + r'_2$, such as the  $41d + 49s \rightarrow 42p + 49p$ almost resonant reaction in rubidium  (see figure \ref{fig:Heuvel}).
The total hamiltonian $H=H_{1+2}$ would contain the non interacting hamiltonian $H_1$ and $H_2$ and   hamiltonian containing all the near resonant interactions $H_{12}$. The reaction $r'_1 + r_2 \leftrightarrow r_1 + r'_2$ is not included because being out of resonance.  Finally,
\begin{eqnarray}
H_{1+2}& =& \hbar \omega_{r_1} |r_1\rangle \langle r_1| +  \hbar \omega_{r'_1} |r'_1\rangle \langle r'_1| +  \hbar \omega_{r_2} |r_2\rangle \langle r_2  |+
 \hbar \omega_{r'_2} |r'_2\rangle \langle r'_2| + \\
 & & H_{12}^{r_1r'_1;r_2r'_2} + H_{12}^{r'_1 r_1;r'_2 r_2} \label{H_1_and_2}
 \end{eqnarray}
where for instance $H_{12}^{r_1r'_1;r_2r'_2}$ being the hamiltonian given by
Eq. (\ref{H_12_ret}) where $\vec \mu_1$,  being replace by $\vec{\mu}_1^{r_1  r'_1}$, $\vec \mu_2$,  being replace by $\vec{\mu}_2^{r_2  r'_2}$ and $k=\omega_{r_1 r'_1}/c \sim - \omega_{r_2 r'_2}/c$.

\subsection{Migration and Förster vision}

The dipole-dipole coupling allows a pair
of Rydberg atoms to exchange the excitation in the reaction $%
np+ns\leftrightarrow ns+np$. Independently of the external field it is an exactly resonant transition which can lead to a diffusion of the $np$ excitation if
surrounded by $ns$ atoms \cite{2004PhRvA..70c1401M}, it is sometimes calls migration reaction. Analogy can
be made with excitation diffusion of excitons \cite{PhysRevB.51.7655} or
spin glasses systems \cite{castellani2005spin}.
Such coupling necessitates nevertheless to
prepare the atoms in different states.

\ Another configuration based on
transition dipoles is the Förster resonance one where two atoms  exchange internal
energy in dipole-dipole interaction, corresponding to different dipole
transitions.
 As simple illustrative example
(see Figure \ref{fig:Stark}), let us consider the cesium case and the
excitation of the $np$ states in an electric field. 
By Stark effect, the $np$
energy level can be put midway between the energy of the $ns$ and $(n+1)s$
states. So that a reaction of the type $np+np\leftrightarrow ns+(n+1)s$ is
resonant (i.e. does not require any energy input). 
The name of "Förster resonance"
has been given by Thad Walker and Mark Saffman
in the reference\footnote{%
Despite the typographical error in the name \textquotedblright Fö%
rster\textquotedblright\ in the title.} \cite{2005JPhB...38S.309W}
 by
analogy with the FRET (Förster Resonant Energy Transfer) process present in
biology \cite{forster48}. The FRET occurs when the dipolar interaction
between two molecules allows excitation transfer between one molecule to the
other one which then becomes fluorescent. The resonance condition comes form
the fact at a given distance the dipolar interaction match the energy
difference of the two molecular systems opening the reaction.
The terminology "Förster resonance" does not seems to be fully defined but is usually used when the two initial states are the same i.e. for 
 reactions of the type $n \ell + n \ell \leftrightarrow n_1 \ell_1 + n_2 \ell_2$ which are 
  listed for several atoms in reference \cite{2008PhRvA..77c2723W}.
  This choice comes because most experiments use a single atomic species (homonuclear case, $A= B$) excited towards the same Rydberg level.
  However, to keep the spirit of the FRET transfer it is probably better to extend the term "Förster" resonance to any type of resonance which, contrary to the migration reaction, are  not always exactly resonant such in as
the rubidium
$41d + 49s \rightarrow 42p + 49p$ reaction.

\subsection{Frontier of the long range van der Waals interaction between Rydberg atoms}

\begin{figure}[!t]
\includegraphics[width=8.5cm]{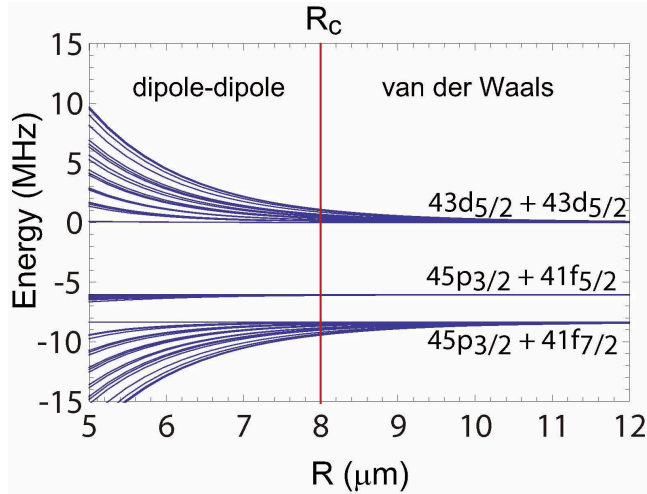}
\caption{Potential energies for the interaction channel $43d_{5/2}+43d_{5/2}%
\rightarrow 45p_{3/2}+41f$ in rubidium Rydberg atoms. The cutoff radius $%
R_c=R_{vdW}$ represents the distance scale for the transition from resonant
dipole-dipole to van der Waals behavior. From \protect\cite%
{2009arXiv0909.4777S}.}
\label{fig.vdWForster}
\end{figure}

The dipole-dipole hamiltonian
 $H_{12}$ 
given by Eq. (\ref{H_12_ret})
 or more simply by  the non retardated Eq. (\ref{int_dipol_dip}) couples a Rybderg pairs of atoms $| r_1 \rangle_A | r_2  \rangle_B = | r_1 , r_2  \rangle$ towards other pairs $| r'_1 , r'_2  \rangle$ creating, in function of the internuclear distance $R$ between the two atoms the potential curves. In presence of electric field   we should use the $|r (F) \rangle$ states, but
to simplify the notations we study the case of the absence of an electric field where there is no permanent dipole moment ($%
\vartheta=0$). 
The
hamiltonian describing the dipole-dipole
interaction, in the atomic bases $| r'_1 , r'_2
\rangle $ and $| r_1 , r_2  \rangle$ is:  
$
\left(%
\begin{array}{cc}
\hbar \Delta & V \\ 
V^\dag & 0%
\end{array}%
\right)%
$
where we have noted the dipole-dipole coupling $V=\langle r_1 , r_2  | H_{12} | r'_1 , r'_2  \rangle$ and $\hbar \Delta = (
E_{r'_1}+E_{r'_2}) - (E_{r_1}+E_{r_2})$ the Förster energy mismatch. $V$ is basically $C_3/R^3 $ with $C_3$ being $e^2
R_{n_1 \ell_1}^{n'_1 \ell'_1 } R_{n_2 \ell_2}^{n'_2 \ell'_2}$ times a coefficient (called $\sqrt 
\mathcal{D}$ in \cite{2008PhRvA..77c2723W}) depending on angular
(Clebsch-Gordan) coefficients. 
 We do not resolve the $m_j$ projection so $V$, or $C_3$, is an operator in this subspace. The same block matrix representation can be used in Eq. (\ref{eq_champ}), and results are generally valid using this block representation.
We note $| \widetilde{ r_1 , r_2} \rangle $ and $|  \widetilde{ r'_1 , r'_2 } \rangle$
 the new eigenstates.
 As an example we show in Fig.~\ref{fig.vdWForster} the $43d_{5/2}+43d_{5/2}$
rubidium potential curves. The two ($j=5/2,7/2$) nearby $f$-states interact
strongly with the $43d_{5/2}$ state. There are states with
extremely small $C_3$, resulting in eigenstates that are nearly flat in
energy and that would make difficult any blockade experiment.

 The 
$r_1+ r_2$ energy shift, which constitutes the $R$-dependent potential curves between the atoms, is given by the equation $\hbar \widetilde{ \Delta }  (R) = \frac{%
\hbar \Delta}{ 2}-\mathrm{sign}(\hbar \Delta )\sqrt{\frac{\hbar \Delta}{4}^2+\frac{C_3^2%
}{ R^6} } $ where $\frac{C_3^2}{R^6} 
V^\dagger V=$ \cite{2008PhRvA..77c2723W}.
 Finally, the component of the new eigenstate $| \widetilde{ r_1 , r_2} \rangle$ on the states $
 |  r_1 , r_2 \rangle $ and $
 |  r'_1 , r'_2 \rangle $ are in ratio $\frac{V}{ \hbar \widetilde{ \Delta} -\hbar \Delta}  $.

We can define a cross-over distance, sometimes called van der Waals radius, $%
R_{vdW}$ via $\hbar \Delta =C_3/R_{vdW}^3$.  $R_{vdW}$ denotes the region where the
energies transition from the van der Waals to the resonant dipole-dipole form.

At small
distances, $R\ll R_{vdW}$  the 
energy shift verifies $\hbar \widetilde \Delta \approx V = C_3/R^3$ which is the largest possible interaction energy between
two Rydberg atoms. 
For instance, in the cesium case
when the two atoms are in
state $|r_1,r_2 \rangle = |np,np\rangle $, they interact strongly
with the symmetric state
$( | r'_1 , r'_2 \rangle  = ( |ns,(n+1)s \rangle +  | (n+1)s , ns \rangle)/\sqrt{2}$ and not with the antisymmetric one $(|ns,(n+1)s \rangle - | (n+1)s , ns \rangle)/\sqrt{2}$. 
This which leads to symmetric
energy shifts E = $\pm C_{3}/R^{3}$, corresponding to the symmetric and
antisymmetric state $$[|np,np\rangle \pm \left( \left\vert
ns,(n+1)s\right\rangle +\left\vert (n+1)s,ns\right\rangle \right) /\sqrt{2}]/%
\sqrt{2}.$$ It is interesting to notice that the two atoms are naturally
entangled in the  Hund's base state
$(|r_1,r_2 \rangle + |r'_1,r'_2 \rangle)/\sqrt{2}$
by the Förster coupling. This has to be compare to the entanglement created by
 the migration reaction $r_1+r_2 \rightarrow r_2 + r_1$ in the
  homonuclear case
 where the best base to
   treat the problem is $( | r_1 , r_2 \rangle \pm  | r_2  r_1 \rangle)/\sqrt{2}$.

 At large distances $R\gg R_{vdW}$, 
the perturbated levels $r'_1 + r'_2 $ are far away (in energy compare to the coupling $V$) of the $ r_1 + r_2$ ones.
When only one
nearby energy level dominates, the energy shift expression becomes $\Delta E \approx 
\frac{(C_3/R^3 )^2}{ \hbar \Delta} = C_6/R^6$. When no single $| r'_1  r'_2 \rangle$ state dominates
the energy level shift of the $r_1,r_2$ state, this shift can still be calculated by using the second order
perturbation theory: $\hbar \widetilde{ \Delta}  = \sum_{r'_1,r'_2} \frac{| \langle r_1 r_2 | H_{12}| r'_1  r'_2 \rangle|^2}{(
E_{r'_1}+E_{r'_2}) - (E_{r_1}+E_{r_2}) }  $.

Concerning the nomenclature some texts mean by van der Waals only the
attractive forces and then sometimes distinguish van der Waals-Keesom, van
der Waals-Debye, and van der Waals-London; others used "van der Waals" terms
to design any kind of interactions repulsive or attractive ones. In this
text we shall distinguish between effect calculated by first order
perturbation theory and effect calculated using second order perturbation
theory. 

The first order perturbation theory, which creates interaction
depening as $1/R^3$ where $R$ is the interatomic distance, can be used when
permanent dipole moment (i.e. Keesom's type) or near resonant transition dipole moment (Fö%
rster's type) exists. 
 This is the case for states $n \ell + n' (\ell \pm 1)$ state
 in homonuclear atoms ($A=B$). Because
of the $A(ns)+B(np) \rightarrow A(np)+ B(ns)$ resonant migration transition  $ns+np$ (i.e. one atom in $n s$ level and the other in $np$ level) present a 
 $1/R^3$ potential curve. This is the levels  used
 in light-assisted collisions or photoassociation 
\cite{1999RvMP...71....1W,2006RvMP...78..483J}.

 The second order perturbation theory is used in other cases, i.e.
when only induced dipole (London's type) moment are considered. It is thus
mainly used in zero electric field, and the interaction energy has a $1/R^6$
dependence as for the $ns+ ns$ potential curves. 
 We shall not consider here more complex interactions, i.e.
higher order terms in the asymptotic expansion, such as quadrupolar one,
because they are usually negligible. For instance interaction in $np+np$
state  is of the $C_5/R^5 + C_6/R^6$
form. However the $C_5/R^5$ quadrupolar term is usually small compare to the van der
Waals $C_6/R^6$ one \cite{2008PhRvA..77c2723W}.

As the dipoles scales as $n^{2}$ the $C_{3}$ coefficient present a rapid $%
n^{4}$ scaling, and because  $\hbar \Delta$ scales as the level spacing i.e. as $n^{-3}$, the $%
C_{6}$ coefficient present a rapid $n^{11}$ scaling.

\section{Conditional and collective laser Rydberg-excitation in the two-atoms blockade regime}
\label{sec_dip_two_at}

Interaction  between cold
atoms have been discussed several times in the context of quantum computation \cite%
{2000PhRvA..61f2309B,2000PhRvL..85.2208J,2001PhRvL..87c7901L,2005JPhB...38S.421R}. Since 2001, a large interest for the Rydberg atoms arise due to two
theoretical propositions by Zoller's group \cite%
{2000PhRvL..85.2208J,2001PhRvL..87c7901L} intended to realize a fast quantum
gate using ultracold atoms excited in Rydberg states. 
 Indeed, the basics of
a quantum gate is the control of a (quantum-)bit by another. This is exactly
what was proposed in these references, and is described on the figure \ref%
{fig:phasegate}: the Rydberg excitation of one atom depends of the
excitation of the first one due to the fact that the energy level is
affected by the dipole-dipole interaction: this is the so called "dipole
blockade" effect. Before considering the laser Rydberg-excitation of a large ensemble of cold
atoms, we would like to introduce  the interaction of two Rydberg atoms
under a laser field excitation. 

We would then described briefly experiments that have performed by
Antoine Browaeys and Philippe Grangier's group (Institut d'Optique Graduate
School, Palaiseau, France), in collaboration with our group, and by Mark Saffman and Thad Walker's group
(University of Wisconsin, USA). Both group
uses very small dipole trap to initially confine small number, ultimately
single atoms \cite%
{2001Natur.411.1024S,2006Natur.440..779B}. The two
experimental setups are similar (even if the dipole trap is smaller in the
French setup ensuring single atom trapping) and the distance between the two
traps can typically be varied between $3$ and $20 \,\mu$m. Several Rydberg
excitation state have been used $43d_{5/2},58d_{3/2}$, for their near
resonant Förster properties at zero field, or $79d_{5/2}, 90d_{5/2},
97d_{5/2}$ for their stronger and stronger van der Waals interaction
strength \cite{2009arXiv0909.4777S}.

\subsection{Principle of the Rydberg blockade}

\begin{figure}[h!]
\centering
\resizebox{ 0.7\textwidth}{!}{
		\rotatebox{-90}{
		\includegraphics*[103mm,22mm][203mm,277mm]{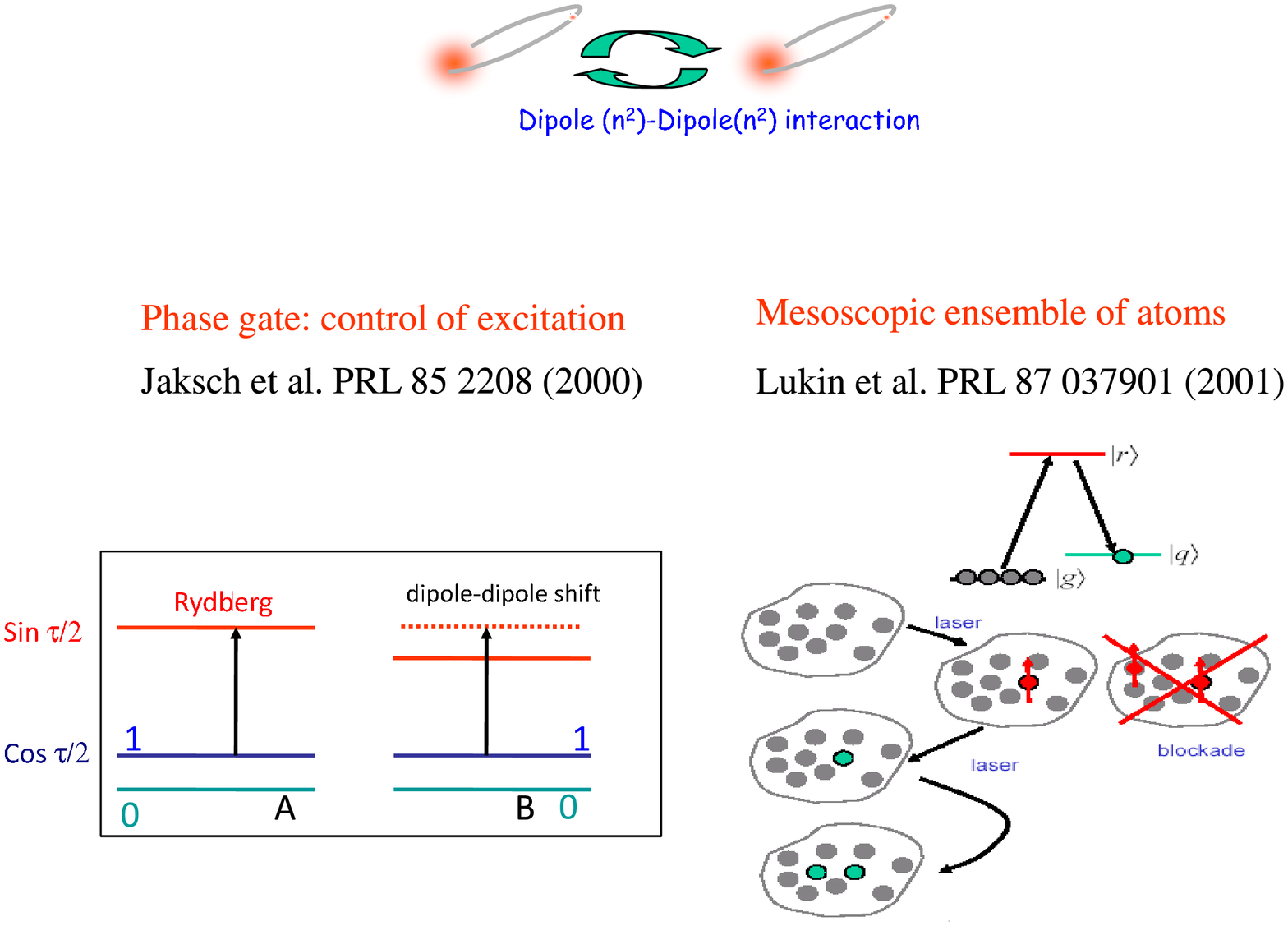}
		}}
\caption{Application of the dipole blockade between two atoms.
Left: the shift created by the dipole-dipole interaction allow to control
the excitation of the second atom $B$ by the first one $A$ by preventing its
excitation when the first atom $A$ has been excited. Right: when generalized
to a larger atomic sample, the shift created by the dipole-dipole
interaction allow a single excitation creating a Fock state. The short lived
Rydberg Fock state can then be transfer by laser to a more stable one.}
\label{fig:phasegate}
\end{figure}

In the dipole
blockade \cite{2001PhRvL..87c7901L} of the laser excitation of cold Rydberg
atoms the ground state $|g\rangle$ and the Rydberg state $|r\rangle$ of
each atom are coupled by a laser with a Rabi frequency $\Omega$. When one
atom is excited in a Rydberg state, the dipole-dipole interaction shifts the
resonance and prevent the laser excitation of the second as explained also
in figure \ref{fig:bloc_browaeyes} a).
More precisely, when the two atoms are in state $|r,r\rangle$, they interact
strongly which leads to symmetric energy shifts $E = \pm C_3/R^3$. When the interaction energy becomes larger than both
the Rabi frequency and the laser resolution, the laser is out of resonance
with the transition coupling the singly with doubly excited state, and only
one atom at a time can be transferred to the Rydberg state.

\subsection{Many photon laser  excitation of two atoms}

\begin{figure}[h!]
\centering
\resizebox{0.7\textwidth}{!}{
		\rotatebox{-90}{
		\includegraphics*[25mm,67mm][191mm,223mm]{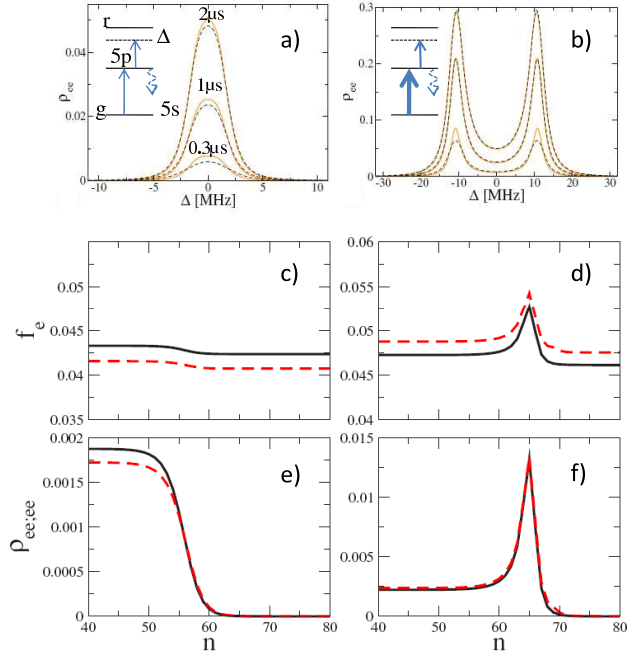}
		}}
\caption{Population of the Rydberg level for the rubidium two-step
excitation. Solid line is the result of the Rate Equation and dashed line of
the full Optical Bloch Equation. Each column shows
results with constant Rabi frequency, but the second column describes
results when an higher laser intensity is used for the first excitation
step, as schematically indicated in the inset of a) and b).
a) and b) display the Rydberg excitation for a single atom (for different pulse length of $%
0.3$, $1.0$ and $2.0\,\protect\mu$s. c)-f) results for two interacting atoms
case (scaled to the case of two $n=48$ (s) atoms separated by $5\,\protect%
\mu $m). $\protect\rho_{ee}$ is the population of the Rydberg level, $f_e$
is the fraction of excited atoms, $\protect\rho_{ee,ee}$ is the probability
that both atoms are in the Rydberg state. The blockade effect is visible for 
$n>60$ in (e), and the anti-blockade is visible for $n\sim 65$ in (f). Adapted from 
\protect\cite{2007PhRvA..76a3413A}.}
\label{fig:Rost}
\end{figure}

The ground state - Rydberg transition is usually in the UV region so experimentalists  usually prefer to excite Rydberg state via an intermediate level.  Before describing the experiments, it is important to know if a more accurate treatment than "forgetting" about this intermediate state,  to recover the previous 2 level  ($g,r$) picture, is  needed or not.
 This procedure has been studied in
detail by the Dresden group \cite{2007PhRvA..76a3413A} in the rubidium $g=5s \rightarrow 5p \rightarrow r$ scheme. The relaxation
of the populations and of coherences have to be taken into account using Optical Bloch Equations. The authors show, see
Figure \ref{fig:Rost}, that
if it is possible to adiabatically eliminate the coherence in the equations to
reach standard two level rate equation, the
adiabatic elimination of the intermediate state has to be done with care. For
instance, when the Autler-Townes splitting, of the intermediate
step, are equal to the dipole-dipole Rydberg shift this increases the Rydberg excitation probability ($n\sim
65$ case in Figure \ref{fig:Rost} f). 
as shown for $n\sim 65$ in Figure \ref%
{fig:Rost} f). This phenomenum has been called by the authors an
"antiblockade" effect because it leads to an non so efficient blockade
mechanism.
This anti-blockade effect can not be predicted using simply the ($g,r$) picture but is easily understand by coupling  (dressing) by laser the ground state with the intermediate level.
The anti-blockade arise when this Autler-Townes  state is put into resonance with the shifted Rydberg-Rydberg
level.

\subsection{Conditional excitation of a single Rydberg atoms}

We are going to treat theoretically two
different cases: the conditional Rydberg excitation of one atom  in presence a
close neighbor Rydberg atom and the collective Rydberg excitation of a pair
of atoms. The two situations refer to two different experiments: the conditional one addresses individually each atom by separate lasers, whereas the collective one addresses both atoms by the same laser.
We treat here the most general case of a non-resonant Förster resonance.
The total hamiltonian is $H_{1+2}$, cf Eq. (\ref{H_1_and_2}), added to the laser interaction hamiltonian:  $-\vec \mu_1.\vec E_L \cos (\omega_L t- \vec k.\vec R_1)$ for the atom $1$ and $-\vec \mu_2.\vec E_L \cos (\omega_L t- \vec k.\vec R_2)$ for the atom $2$.

 For a conditional excitation,
we consider a one-photon excitation driving.   The ground state $(g_2)$ of the second atom toward Rydberg state $|g_2\rangle \rightarrow |r_2\rangle$ transition with a  Rabi frequency $\Omega = \langle g_2 |-\vec \mu_2.\vec E_L | r_2 \rangle/\hbar $  and a laser detuning  $\delta=\omega_L - \omega_{g_2 r_2} $. 
The first atom is previously excited in a Rydberg state $r_1$. We have
to consider the transition $|r_1 ,g_2\rangle \leftrightarrow 
| r_1,r_2\rangle $ and the Förster coupling $| r_1,r_2\rangle \leftrightarrow | r'_1,r'_2\rangle$ (strength $V$).   We 
 choose to decompose the wavefunction as:
\begin{equation*}
\psi \left( t\right) =a_{g}(t)  \left\vert r_1,g_2\right\rangle
+ e^{i \vec k.\vec R_2} e^{- i \omega_L t} \left[ a_{r}(t)   \left\vert r_1,r_2\right\rangle + a_{F}(t)  |r'_1 r'_2 \rangle \right].
\end{equation*}%
 The evolution of the system is given by (with energy origin for the state ($\left\vert
r_1,r_2\right\rangle$):
\begin{eqnarray*}
i\frac{da_{g}}{dt} &=&\delta a_{g}+\frac{\Omega }{2}a_{r} \\
i\frac{da_{r}}{dt} &=&\frac{\Omega^* }{2}a_{g} + (V^*/\hbar) a_{F} \\
i\frac{da_{F}}{dt} &=&(V/\hbar) a_{r} + \Delta a_{F}
\end{eqnarray*}%
Where we have performed the rotation wave approximation consisting to neglect the non resonant terms.
We are not solving here the equation, but 
its resolution  is easy, especially in the Förster resonance case of $\Delta=0$, if using the $| \widetilde{r_1,r_2}\rangle$ base.

\begin{figure}[t]
\includegraphics[bb=14mm 50mm 92mm 274mm,clip,angle=-90,width=10cm]{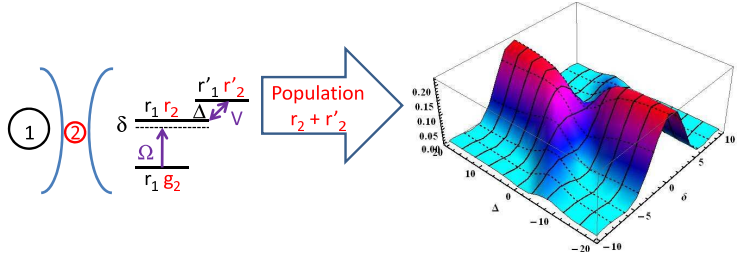}
\caption{Calculations of the conditional excitation versus
the laser-excitation, $\delta $, and the Förster resonances detuning $\Delta 
$. The calculation is performed at small time $ \Omega t=1$ and in a strong coupling ($V = 3 \hbar \Omega$) regime. At Förster resonance, $\Delta =0$, we observe a  blockade
effect of the excitation at resonance $\delta =0$.}
\label{fig:exc_cond2}
\end{figure}

 The figure \ref{fig:exc_cond2} shows in a perturbative regime for the excitation the
probability of excitation versus laser excitation resonance, $\delta $, and
the Förster resonance, $\Delta $. 
At Förster resonance, $\Delta =0$, we observe a  blockade
effect of the excitation at laser resonance $\delta =0$. This corresponding to the lift of
the degeneracy between $\left\vert r_1 r_2 \right\rangle$ and $\left\vert r'_1 r'_2 \right\rangle $ state due to the dipole-dipole coupling. 
When scanning the laser detuning $\delta$ this leads also to a
broadening, or even a splitting in our strong coupling case, of the Förster
resonance.

\begin{figure}[t]
\includegraphics[bb=12mm 52mm 99mm 274mm,clip,angle=-90,width=10cm]{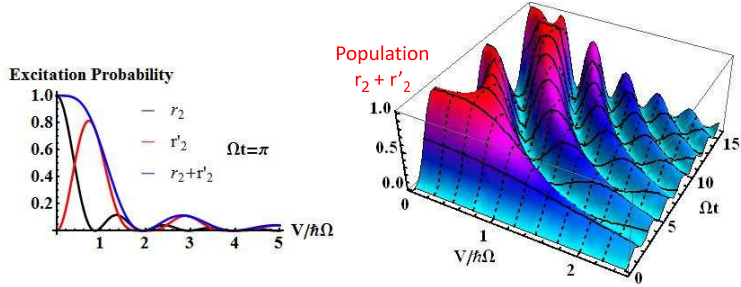}
\caption{Conditional excitation of a second atom into a Rydberg state $|r_2\rangle$ when a neighboring atom is already in Rydberg state. Same notations as in figure \ref{fig:exc_cond2}. Excitation probability  in function of the strength of the 
dipole-dipole interaction $V $, 
of the interaction time $t$ (in Rabi $\Omega$-frequency units). Rabi oscillation are visible in the time evolution. The blockade of
the excitation is obtained as soon as $V \gg \hbar \Omega $.}
\label{fig:exc_cond1}
\end{figure}
The figure \ref{fig:exc_cond1} shows, in a more detail way,
the probability for the Rydberg excitation of a second atom versus the Fö%
rster coupling
 for a square pulse excitation.
As visible in the 
 case of  a $\pi$ excitation ($\Omega t=\pi $), for intermediate coupling
($V\simeq \Omega $)
 a large amount of population are in the Förster Rydberg states $r'_1$ and $r'_2$.

\subsection{Experimental realization of conditional Rydberg excitation}

As mentioned, previously, propositions to use pairwise controlled
dipole-dipole interactions \cite{2000PhRvA..61f2309B} as a very efficient
realization of a quantum logic gate or to entangle neutral atoms has pushed
experimentalist to realize the gedanken experiment proposed in reference 
\cite{2001PhRvL..87c7901L}.

\begin{figure}[h!]
\centering
\resizebox{0.7\textwidth}{!}{
				\includegraphics*[0mm,57mm][91mm,117mm]{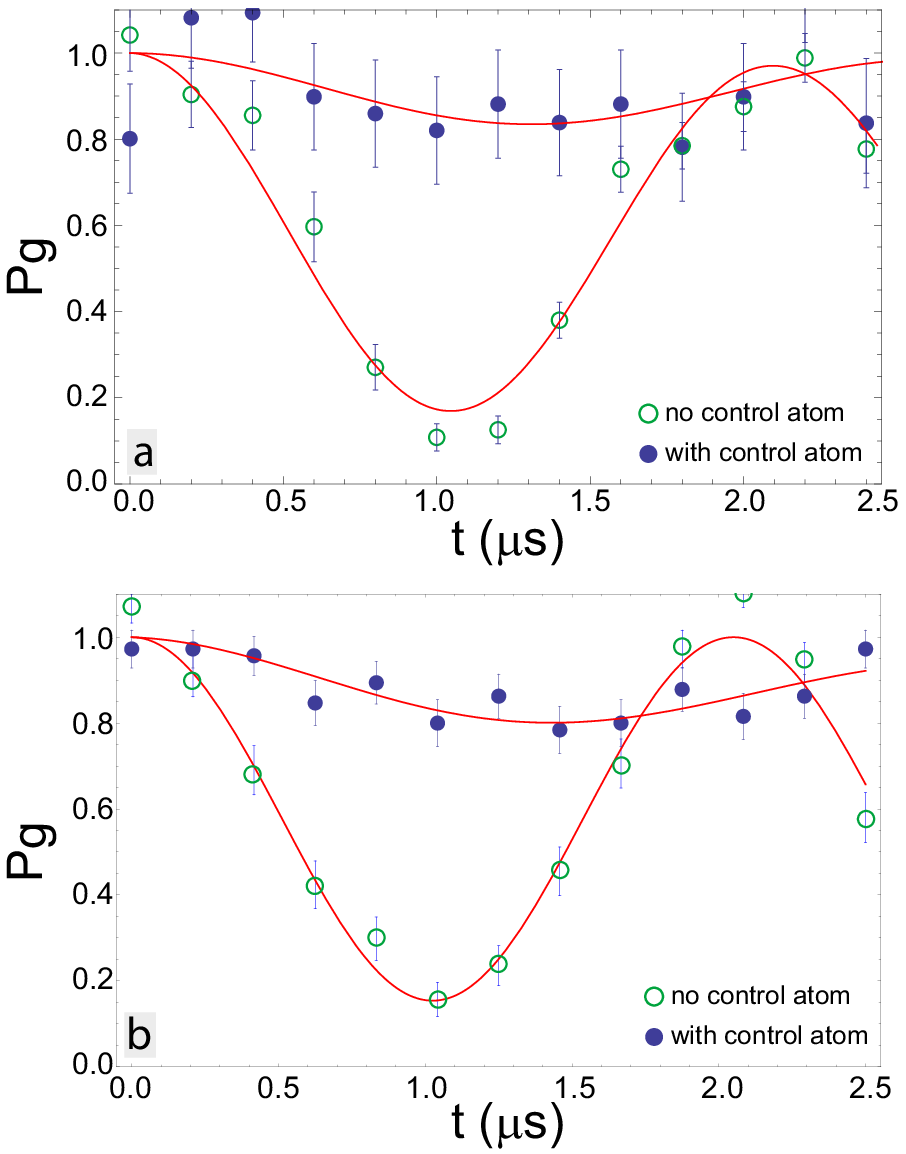}
				}
\caption{Single atom Rabi oscillation and evidence for blockade of the
Rydberg excitation when a second atom is present. Experimental data for
Rydberg excitation of the target atom with and without a second (control)
atom present. From \protect\cite{2009NatPh...5..110U}}
\label{fig:bloc_Saffman}
\end{figure}

Rabi oscillation (previously observed by M. Weidemüller's group in an larger
atomic sample \cite{2008NJPh...10d5026R}) has first been observed, by the
Wisconsin team, in laser excitation of a single trapped atom. This has
indicated that the system is not submitted to strong decoherence effects and
has opened the way to further experiments \cite{2008PhRvL.100k3003J}. The
dipole blockade of the laser excitation, for a pair of individually trapped
atoms, has indeed been observed recently by the two experimental groups \cite%
{2009NatPh...5..110U,2009NatPh...5..115G}. Two atoms are confined in two
independent optical dipole traps, which are usually turned off during the
excitation to avoid an extra light-shift. A successful excitation of an atom
to the Rydberg state is detected through the loss of the atom when the
dipole trap is turned back on, as atoms in the Rydberg state are not trapped
in the tweezers. One of the key result is that single Rydberg-excited
rubidium atom blocks excitation of a second atom located more than $10\, \mu$%
m away (this so called blockade radius depends on the choice of the Rydberg state) has
shown for instance in figure \ref{fig:bloc_Saffman}.

\subsection{Experimental realization of collective excitation of a pair of Rydberg atoms}

The collective two-atom behavior has also been experimentally demonstrated
with the excitation of an entangled state between the ground and Rydberg
levels \cite{2009NatPh...5..115G}. In this experiment both atoms are
simultaneously illuminated by the excitation laser beams. 

\begin{figure}[h!]
\centering
\resizebox{0.7\textwidth}{!}{
		\rotatebox{-90}{
		\includegraphics*[40mm,25mm][170mm,130mm]{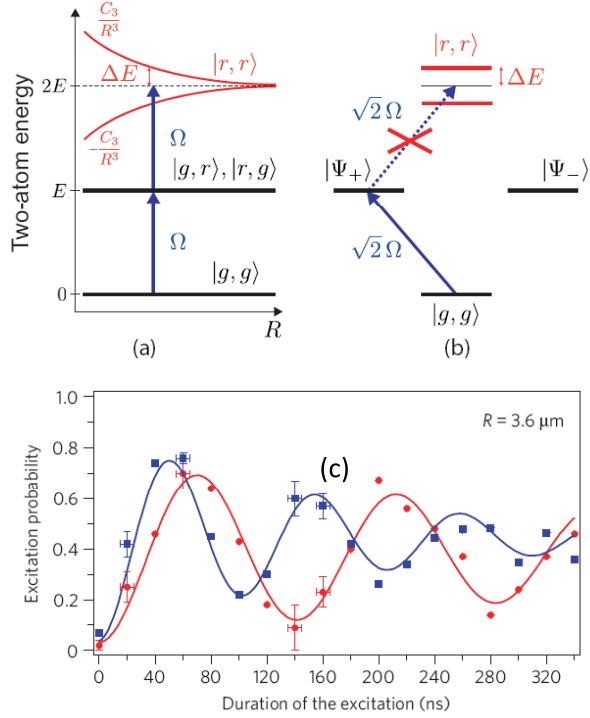}
		}
		}
\caption{Principle and experimental realization of the Rydberg blockade. (a)
Principle of the blockade between two atoms, in the regime of conditional
excitation. When both atoms are simultaneously excited in the blockade
regime, the symmetrical state $| \Psi_+ \rangle$, described in the text, is
only coupled to the ground state $|g,g \rangle$ with a strength $\protect%
\sqrt{2} \Omega$ while the state $| \Psi_- \rangle$ is not coupled by the
laser to the states $|g,g \rangle$ and $|r,r \rangle$. c) Collective
excitation of the two atoms separated by $3.6\,\protect\mu$m. The circles
represent the probability to excite one atom when the second atom is absent.
The squares represent the probability to excite only one atom when the two
atoms are trapped and are exposed to the same excitation pulse. From 
\protect\cite{2009NatPh...5..115G}}
\label{fig:bloc_browaeyes}
\end{figure}

The principle is indicated in
Figure \ref{fig:bloc_browaeyes} (a), when both atoms are simultaneously excited.
A fundamental consequence of the blockade is that as any of the two atoms
can carry the excitation, they end up in the entangled state $|\Psi_+
\rangle =  \frac{1}{\sqrt{2}} \left( |g,r\rangle e^{ik R_2} + |r,g\rangle
e^{ik R_1} \right) $, where $R_1$ and $R_2$ are the position of the two
atoms, and $k$ the wavevector of the excitation. The coupling from the
two-atom state $|g,g\rangle$ toward the state $|\Psi_+ \rangle$ is $\sqrt{2}
\Omega$, while the state $|\Psi_- \rangle = \frac{1}{\sqrt{2}} \left(
|g,r\rangle e^{ik R_2} - |r,g\rangle e^{ik R_1} \right) $ is not coupled
with the ground state. In the blockade regime, the two atoms are therefore
described by an effective two-level system involving collective states $%
|g,g\rangle$ toward $|\Psi_+ \rangle $ with a Rabi frequency $\sqrt{2}
\Omega $.

The figure \ref%
{fig:bloc_browaeyes} (b) shows the experimental result where the Rydberg excitation is
applied either to a single atom or to two neighboring atoms at a distance of 
$3.6\,\mu$m. The probability to excite both atoms in the Rydberg state is
suppressed, as it is expected in the blockade regime. The probability to
excite only one of the two atoms as a function of the duration of the
excitation, can be compared with the probability to excite one atom inside a
given trap when the other trap is empty. The two probabilities oscillate
with different frequencies in ratio $1.38 \pm 0.03$ close to the expected
value of $\sqrt{2}$. This result is the signature of the collective behavior
of the two atoms and has been used to create the entanglement of two
hyperfine ground state atoms by laser inducing the decay of the Rydberg
state. The entangled state, reached with a fidelity of $0.75$, is generated
in only $200\,$ns using pulsed two-photon excitation and has been quantify
by applying global Raman rotations on both atoms \cite{2010PhRvL.104a0502W}.
Finally the first demonstration of a CNOT gate, with fidelity of $0.73$,
using neutral atoms have been achieved in Wisconsin \cite%
{2010PhRvL.104a0503I}. 

Both the Palaiseau and Wisconsin experiments suffer from  atoms losses
during the gate operation and results have to be improved to reach
the high fidelity results obtained with trapped ions. However these exciting recent experiments, pave the
way for demonstrating more complex operations with several qubits \cite%
{2009arXiv0909.4777S}. Finally, such clean experiments performed in sample
containing only few atoms have brought better understanding of the
collective behavior as well as on the decoherence processes occurring in
Rydberg assembly. Such, small atom number experiments are ideal to study of
a Rydberg sample.
Because a recent  review article cover these subjects we shall not focus more on it here \cite{2009arXiv0909.4777S}. We simply give a simple theory able to describe the system

\subsection{Theory for collective excitation of a pair of Rydberg atoms}

\begin{figure}[t]
\includegraphics[bb=30mm 26mm 166mm 279mm,clip,angle=-90,width=10cm]{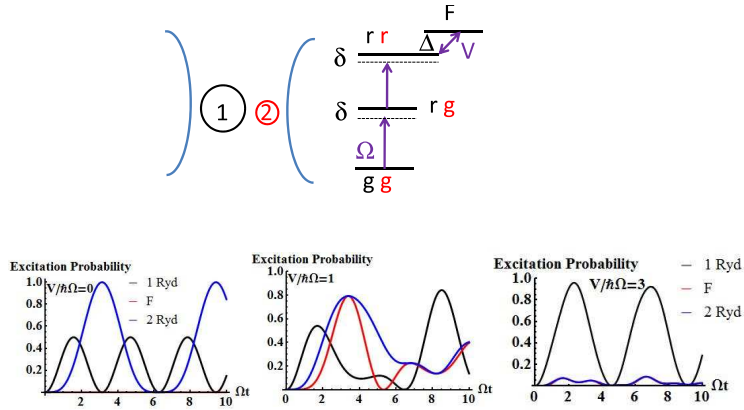}
\caption{Collective excitation of  two atoms, i.e. creation of a single but delocalized Rydberg excitation. 
The laser, of single atom Rabi frequency $\Omega$, is tuned at the resonance $\protect\delta =0$, and the Rydberg state is energy shifted by a resonant Förster dipole-dipole interaction of strength $V = \langle F | H_{12} | r,r\rangle$ equals to $0,1$ and $3$ times $\hbar \Omega$ . We present the probability that one atoms (number $1$ or $2$) is excited in the Rydberg state $|r\rangle$, that two atoms are excited but in the Förster state $|F\rangle$ and that the two atoms are excited in any Rydberg states $|r,r\rangle$ or $|F\rangle$.
The blockade occurs when the Rydberg state is  shifted more than the Rabi frequency i.e. for $V\gg \hbar \Omega$. Rabi oscillation are visible and, in the blockade regime, with a $\sqrt{2}$ speed enhancement for the collective excitation (single delocalized Rydberg excitation)  compare to the single atom one. }
\label{fig:exc_coll}
\end{figure}

The case of the collective excitation of the pair of atoms is a little bit
different from the conditional excitation.
\ Indeed, to realize a collective excitation both atoms should be undiscernable for the laser excitation, we call $|r\rangle$ the target Rydberg level and $|F\rangle $ the Förster state coupled to $|r, r\rangle$, for instance in the experiment described in figure \ref{fig:bloc_browaeyes}, $|r\rangle = |58 d \rangle$ and $|F\rangle  = \left( \left\vert 60p,56f \right\rangle +\left\vert
56f, 60p \right\rangle \right) $.
A good basis to study the process is the Dicke
states: 
\begin{align}
	\left\vert g,g\right\rangle & & \nonumber \\
	\left\vert \Psi_+\right\rangle =\frac{1}{\sqrt{2}} \left( |g,r\rangle e^{ik R_2} + |r,g\rangle
e^{ik R_1} \right) & & \nonumber \\
 \left\vert \Psi_-\right\rangle =\frac{1}{\sqrt{2}} \left( |g,r\rangle e^{ik R_2} - |r,g\rangle
e^{ik R_1} \right) & & \nonumber \\
 \left\vert r,r\right\rangle & & \nonumber 
\end{align}
 with the Förster state
$\left\vert F \right\rangle $  added. 
We note that there is no laser coupling between the ground state $\left\vert
g,g\right\rangle $ of the pair of atoms and the antisymmetrical state $%
\left\vert -\right\rangle $. So
the
wave-function for the pair of atoms is written as %
\begin{equation*}
\psi (t) =b_{g}(t)  
 \left\vert g,g\right\rangle
+b_{+}(t) e^{- i \omega_L t} \left\vert \Psi_+ \right\rangle  + e^{- 2 i \omega_L t} e^{ik R_1} e^{ik R_2} \left[ b_{r}(t)  \left\vert r , r\right\rangle
+b_{F}(t)  \left\vert F \right\rangle \right]
\end{equation*}%
The equations for the laser excitation are%
\begin{eqnarray*}
i\frac{db_{g}}{dt} &=&2\delta b_{g}+\frac{\Omega \sqrt{2}}{2}b_{+} \\
i\frac{db_{+}}{dt} &=&\frac{\Omega^* \sqrt{2}}{2}b_{g} + \delta b_{+}+\frac{\Omega \sqrt{2}}{2} b_{r}\\
i\frac{db_{r}}{dt} &=&\frac{\Omega^* \sqrt{2}}{2}  b_{+} + (V^*/\hbar) b_{F} \\
i\frac{db_{F}}{dt} &=&(V/\hbar)  b_{r} + \Delta b_{F}\end{eqnarray*}%
The coupling with the symmetrical one $%
\left\vert +\right\rangle $ is the Rabi frequency, $\Omega $, for the
excitation of a single atom multiplied by a factor $\sqrt{2}$. The figure \ref{fig:exc_coll}
shows at resonance laser $\protect\delta =0$, and for a resonant Förster configuration $\Delta=0$, the transfer of populations for a pair of atoms in the Rydberg states
out  and in the blockade regime. We have represented: $|b_+|^2$ which represent the probability that one atoms is excited, $|b_r|^2+|b_F|^2$ that two atoms are excited  and $|b_F|^2$ that the population is in the Förster state.

\section{Dipole-dipole interaction in a many atoms gas sample}

\label{sec_dip_many_at}

The dipole blockade effect can in principle  be generalized to an assembly of $N$ atoms. However,
the energy shift created by the dipole-dipole interactions creates, in a
randomly distributed sample, a band of energy level and not anymore simply two levels as for the two atoms case.
In simple case, such as in van der Waals one, the energy levels mainly
depend on the number of excited Rydberg states and the principle of the blockade can be easily extended.
 However, in pure dipolar ($%
1/R^3$) cases the band structure can be very complex. Indeed, even in a
simple three atomic system, T. Pohl and P. R. Berman \cite%
{2009PhRvL.102a3004P} have shown that a noninteracting (so called zero
energy) state exist even if strong two-body interaction exists. Existence of
such zero states is very interesting to study but  is problematic for blockade experiments because they could
be accessed by many-photon excitation even when the two-photon excitation is
well blockaded. 

We are now going to describe experiments that have tried, and succeed, to observe the dipole- blockade in a many atoms system. In a simple picture,
if the volume of the laser excitation is small enough, no second atom can be
excited after the Rydberg excitation of a first one, producing an atomic
ensemble in a singly excited collective state as pictured in Figure \ref%
{fig:phasegate}. Thus, the dipole-dipole interaction between Rydberg atoms
should lead to a limitation of the number of excited atoms i.e. a partial,
or local blockade of the excitation.

Depending on the size of the sample the blockade can be total (single
excitation allowed) or limited. It may be useful to introduce the ``blockade
sphere'' picture \cite{2004PhRvL..93f3001T} which is especially well adapted
to the quasi-isotropic van der Waals interaction. The excitation of a single
Rydberg atom prohibits, via the blockade mechanism, subsequent excitations
of other ground-state atoms within the radius $R_b$ of the blockade sphere.
Since the $N_b$ atoms within the blockade sphere are indistinguishable, they
behave as a ``superatom'' \cite{2006NatPh...2..801V}. The term ``superatom''
is given here by analogy with the superatom in clusters physics which is a
cluster of atoms that exhibits some of the properties of elemental
atom. Here the superatom interacts with the excitation light via a
collective $\sqrt{N_b}$ enhancement of the Rabi frequency \cite%
{2009PhRvA..80c3418S}. The dipole blockade effect implies that only a single
atoms can be excited among $N_b$, into the state $$|\Psi_+ \rangle = \frac{1}{%
\sqrt{N_b}} \sum_i |1(g), \cdots, (i-1)(g),i(r), (i+1)(g),  \cdots, N_b(g)\rangle e^{i \vec
k.\vec R_i} $$, creating a faster (by a factor $\sqrt{N_b}$) Rabi oscillation
of the one atom excitation compare to the Rabi oscillation of a single
isolated atom. This behavior has indeed been observed on many atomic system
by T. Pfau's team \cite{2007PhRvL..99p3601H} before being very clearly
observed in the case of $N_b=2$ by the Palaiseau group \cite%
{2009NatPh...5..115G,2009arXiv0910.0729B}.

\subsection{Saturation of the Van der waals Rydberg excitation}

\begin{figure}[h!]
\centering
\resizebox{0.7\textwidth}{!}{
				\includegraphics*[0mm,0mm][86mm,70mm]{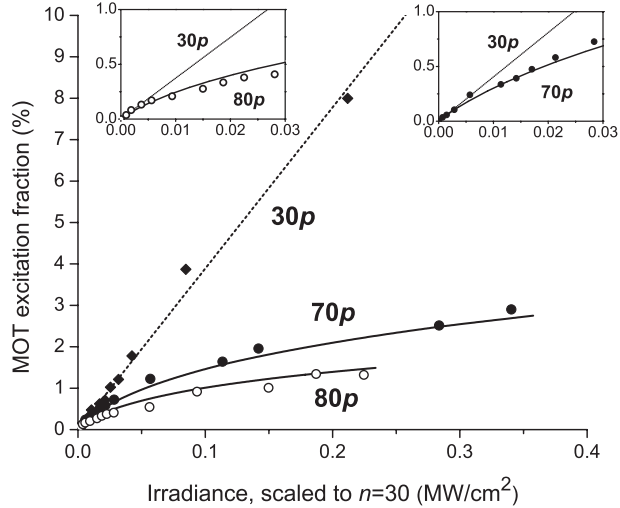}
				} .
\caption{Van der Waals blockade of the excitation. Dependence of the Rydberg
excitation fraction in function of the laser power exciting Rydberg state
for several $n$ principle quantum numbers. For large $n$ the saturation of
the Rydberg excitation clearly indicates a blockade effect. From 
\protect\cite{2004PhRvL..93f3001T}}
\label{fig:vdW_Gould}
\end{figure}

In order to observe the blockade effect the standard pulsed dye laser
spectroscopy is not accurate enough and a spectroscopic accuracy on the
order of the dipolar interaction involved (typically $\sim 10-100\,$MHz) is required. In
2004, P. Gould and E. Eyler's group in Connecticut \cite{2004PhRvL..93f3001T} uses a narrowband pulsed
dye laser to realize the first demonstration of a partial van der Waals
blockade as shown in figure \ref{fig:vdW_Gould}.

However experiments using Continuous Wave (cw) lasers are the more numerous
ones. The first one has been reported by Matthias Weidemüller's group in
2004 (Fribourg, Allemagne) \cite{2004PhRvL..93p3001S}. The main difficulties
of such cw experiments is that the presence of a single ion during the
excitation can stop the excitation and then mimic the effect of the dipole
blockade as probably observed in this first experiment \cite%
{2007PhRvA..76a3413A,Amo08}. Indeed, a single ion creates a spurious field
of $150\,$mV/cm at a distance of $10\,\mu$m which is roughly hundred times
higher than the field created by the dipole of a ($n=50$) Rydberg state.
Such field could easily shifting energy levels by few MHz ($300\,$MHz for $n
= 50$ Rb atoms), i.e. larger than the laser linewidth creating a blockade of
the excitation. Thus, to avoid any formation of ions during the excitation, the duration of the exciting laser light should be
short enough. Another
limitation of the blockade occurs for a broad-band or for a high-intensity
excitation. In such cases, the suppression of the excitation is no longer
expected, since pairs of close atoms can be excited out of resonance.
However, despite all these difficulties, as we shall see several groups
(including Weidemüller's one) have succeeded to observe the Rydberg blockade
of the excitation.

\begin{figure}[h!]
\centering
\resizebox{ 0.7\textwidth}{!}{
		\rotatebox{-90}{
		\includegraphics*[23mm,59mm][97mm,256mm]{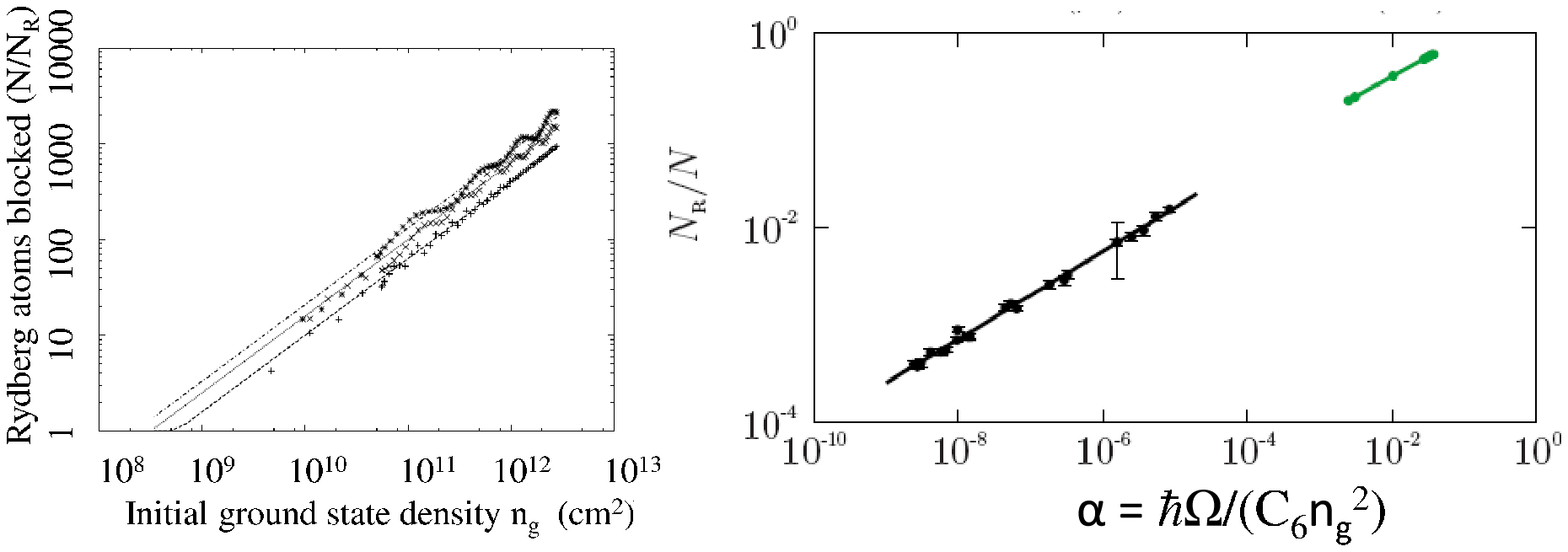}
		}}
\caption{Scaling of the Rydberg number $N_R$ versus the initial atom number $%
N$ (or density $\protect n_g$). Left theory (adapted from \protect\cite%
{2008JPhB...41d5301H}), the three sets are for three different value of the
(single atom) Rabi frequency $\Omega$, the line are the simple law $N_R \sim
\left( \frac{4\protect\pi}{3} \protect n_g \protect\sqrt{\frac{C_6 }{\hbar
\Omega} } \right)^{4/5} \propto \protect\alpha^{-0.4}$ . Right (adapted from 
\protect\cite{2009PhRvA..80c3422L}). Experimental points (black) and
theoretical points (green) with fit $N_R/N \propto \protect\alpha^{0.45}$ for the experiment and $N_R/N \propto \protect\alpha^{0.4}$ for the theory.}
\label{fig:scaling_law}
\end{figure}

As studied by  Francis
Robicheaux \cite{2008JPhB...41d5301H} and Hans Peter Büchler \cite{2008PhRvL.101y0601W}, and recently observed by the
Stuttgart's group\cite{2009PhRvA..80c3422L} (see figure \ref{fig:scaling_law}%
), the number $N_b$ of blocked Rydberg atoms (total number of atoms divided
by the number of excited Rydberg atoms) displays algebraic scaling laws,
with a universal exponent, in function of the initial ground state atomic
density $n_g$. This can be seen as a validation of the blockade radius
picture. Indeed, in the van der Waals regime one simple scaling law is $N_b
\sim \left( \frac{4\pi}{3} n_g \sqrt{\frac{C_6}{\hbar \Omega} }
\right)^{4/5}$. This is derived from the fact that the number of atoms
blocked per excited atom is $N_b \sim n_g \frac{4\pi}{3} {R_b}^3 $ (when the
excited atom blocks all other atoms within the spherical volume of blockade
radius $R_b$) and from $C_6/{R_b}^6 \sim \hbar \Omega \sqrt{N_b}$ because the
blockade occurs when the detuning equals the excitation linewidth (i.e. the
Rabi frequency for sufficiently narrow linewidth laser).

\subsection{Statistic of the Rydberg excitation}

\begin{figure}[h!]
\centering
\resizebox{0.7\textwidth}{!}{
		\rotatebox{-90}{
		\includegraphics*[23mm,30mm][162mm,251mm]{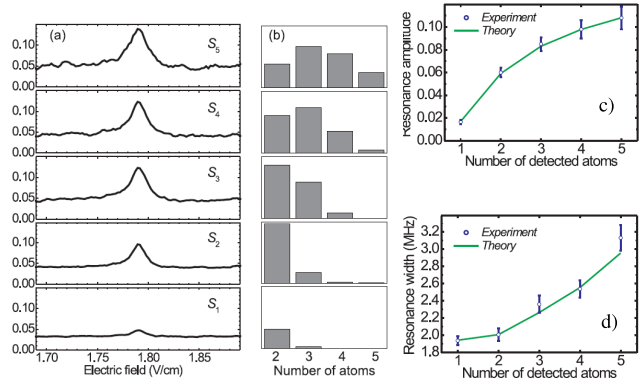}
		}}
\caption{Small atom number Rydberg excitation. a) Experimental spectra of
the Förster reaction in rubidium $37p_{3/2} + 37p_{3/2} \rightarrow
37s_{1/2} + 38s_{1/2}$ for atoms randomly positioned in a ($18\,\protect\mu$%
m)$^3$ cubic volume. b) Theoretical corresponding probability distributions
for the number of actually interacting Rydberg atoms. (c,d) Monte Carlo
Theory and experiment for the resonance amplitude and resonance width. From 
\protect\cite{2010PhRvL.104g3003R}}
\label{fig:ForsterRyabtsev}
\end{figure}

Evidence for the dipole blockade of Rydberg excitation can also examined in
the statistical distributions of the number of Rydberg excitations created
in ensembles of interacting Rydberg atoms. Indeed, in a simple picture, a
fully saturated sample would lead to a well defined number of excited
Rydberg atoms equal to the volume of the sample divided by the blockade
sphere volume. The number of Rydberg atoms should then be well defined i.e.
highly sub-Poissonian. This has been observed first by Georg Raithel's group
(Ann Arbor, Michigan) and confirm later by the Novosibirsk's group \cite%
{2005PhRvL..95y3002L,2007PhRvL..98j9903L,2009PhRvA..79e2504B}. Precise study
of the effect on the finite detection efficiency has been performed \cite%
{2007PhRvA..76a2722R,2007PhRvA..76d9902R,2010PhRvL.104g3003R} allowing to
spectroscopically study few cold rubidium Rydberg atoms confined in a small
laser excitation volume as shown in figure \ref{fig:ForsterRyabtsev}.

\subsection{Probing the band interaction: blockade or antiblockade effects}

 Experimental
investigation of the band of energy level,
created by the dipole-dipole interactions,
 has been first performed by James Martin's group
(Waterloo, Canada) \cite{2004PhRvL..93w3001A} using a radio-frequency probe
transition (see figure \ref{fig:band_Martin}). The author observe broadening
of the band when adding, by RF transfer, strongly interaction Rydberg atoms.
A closely related effect have been observed by T. Gallagher's group, but through direct broadening
of the energy-transfer resonances $np + n^{\prime }s \rightarrow n^{\prime
}s + np$ by introducing an additional Rydberg state $(n^{\prime }+1)p$ which
does not participate directly in the energy-transfer process but is strongly
coupled to the final states $n^{\prime }s$ ($n^{\prime }=33$ in the
experiment) \cite{2004PhRvA..70c1401M}. Since then several group have
studied similar behavior \cite{2009JETP..108..374T} (see figure \ref%
{fig:ForsterRyabtsev} (d)). These experiments indicates that the measured
linewidths are sensitive to the Rydberg atomic density, indicating the
influence of interatomic interactions.

\begin{figure}[h!]
\centering
\resizebox{ 0.7\textwidth}{!}{
		\rotatebox{-90}{
		\includegraphics*[9mm,16mm][190mm,276mm]{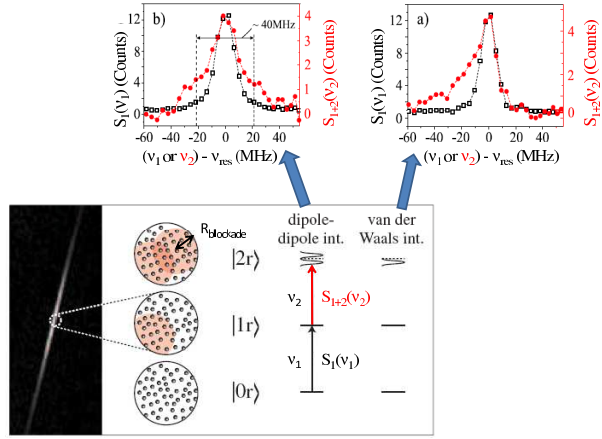}
		}}
\caption{Experimental study of energy levels band spectrum in the van der
Waals and in a pure dipolar interaction cases. The (left) low panel shows a
shadow image of atoms in an optical dipole trap. The right panel shows the
blockade radius $R_b$ as well as an excitation domain with zero, one, or two
interacting Rydberg excitations. Upper figures: spectra for zero applied
field (a) and for an applied field to reach the Förster resonance (b). $S_1(%
\protect\nu_1)$ (black squares; left axis) scan corresponds to the
transition $|0r\rangle \rightarrow |1 r\rangle$. In order to probe the
transition $|1r\rangle \rightarrow |2 r\rangle$, $\protect\nu_1$ is kept at
resonance $\protect\nu_{\mathrm{res}}$ while a second pulses $\protect\nu_2$
is scan and extra signal is recorded in the $S_{1+2}(\protect\nu_2)$ scan
(red circles; right axis). Adapted from \protect\cite{2008PhRvL.100w3201R}.}
\label{fig:two_step_Raithel}
\end{figure}

\begin{figure}[h!]
\caption{Experimental investigation of the broadening of the $45d_{5/2}$
Rydberg energy levels when dipole-dipole interaction (i.e. when $46p_{3/2}$
atoms) is added to the system. The probing of the $45d_{5/2}$ Rydberg energy
levels is performed using a two-photon $45d_{5/2} \rightarrow 46d_{5/2}$
microwave transition. Adapted from \protect\cite{2004PhRvL..93w3001A}.}
\label{fig:band_Martin}\centering
\resizebox{ 0.7\textwidth}{!}{
		\rotatebox{-90}{
		\includegraphics*[90mm,153mm][154mm,266mm]{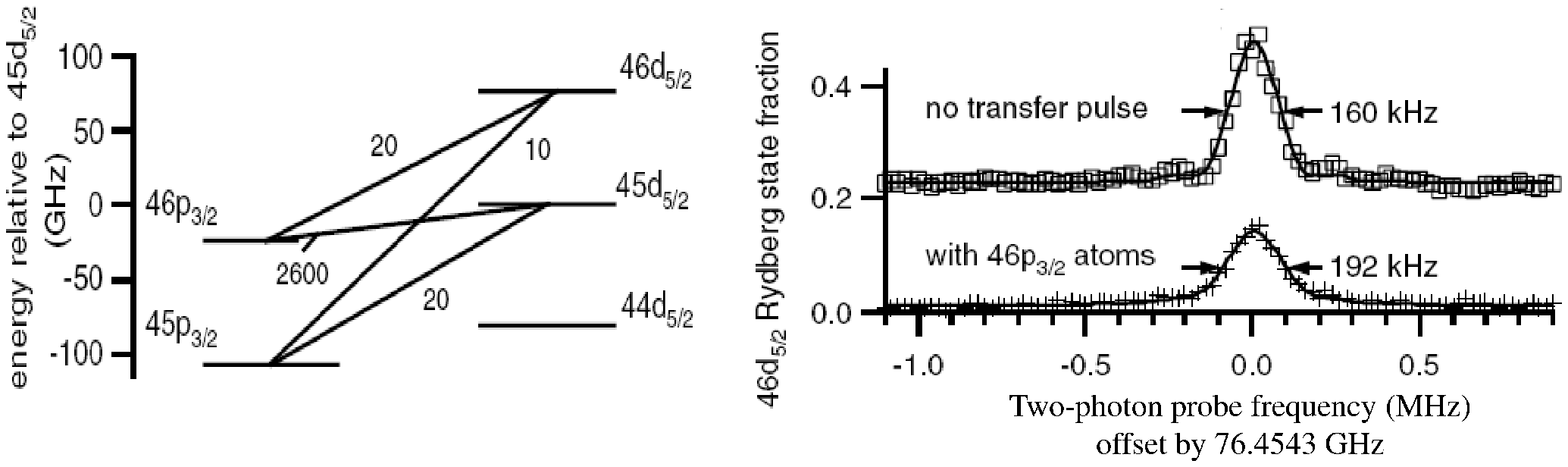}
		}}
\end{figure}

The difference in energy levels band spectrum in the van der
Waals interaction or in a pure dipolar cases has been indeed observed using
cold atoms in an optical dipole trap \cite{2008PhRvL.100w3201R}. The
experiment uses two pairs of independently tunable laser pulses to
spectroscopically probe the spectrum in a double-resonance excitation
scheme. By increasing the magnitude of an applied electric field, the
Rydberg-atom interactions vary from van der Waals (at zero electric field)
to dipole-dipole ($45d_{5/2} + 45d_{5/2} \rightarrow 43f_{7/2} +
47p_{3/2}$) resonant Förster configuration, leading to characteristic signatures in
the measured spectra as shown in figure \ref{fig:two_step_Raithel}. The
zero-field spectrum (a) exhibits a wing on the negative side providing
evidence for a band of two-Rydberg ($|2 r\rangle= |r r\rangle$) excitation showing that
the interactions among $45d_{5/2}$ atoms are negative (attractive) and
primarily van der Waals ($C_6/R^6$) in nature. The red spectrum (b) taken at a field
creating a
resonant Förster configuration,
exhibits symmetric wings, providing evidence for two bands ($\pm C_3/R^3$)
of $|2 r\rangle$ excitation frequencies symmetrically located consistent
with the effect of a Förster resonance. It should be noted that contrary to
the two atoms picture given in the bottom part of the figure no strong
blockade is experimentally observed at resonance. This is due to the angular
averaging effect created by all the binary interactions present ($40$ atoms
are excited in this experiment).

This experiment highlight also that even in presence of a strong dipole interaction  creates a band of energy levels, this should not necessary lead to an efficient
broadening of the direct laser excitation. Indeed, the wings of the band is
produced when two atoms are strongly shifted by interactions and are
therefore not accessible, due to blockade, in a single excitation step, as
shown by the non broadening black curve in figure \ref{fig:two_step_Raithel}%
).

\begin{figure}[h!]
\caption{Observation of the antiblockade effect. Comparison between
calculated (a) and measured (b) $62d$ Rydberg excitation spectra (upper
graphs black) and Penning ionization spectra (lower graphs red) taken at
different time delay. Model and experiments have an Autler-Townes
splitting of $100\,$MHz and a density of the trapped ground state atoms of $%
7\times 10^9\,$cm$^{-3}$.  From \protect\cite{2010PhRvL.104a3001A}.}
\label{fig:antiblokade}\centering
\resizebox{ 0.7\textwidth}{!}{
		\rotatebox{-90}{
		\includegraphics*[42mm,42mm][171mm,257mm]{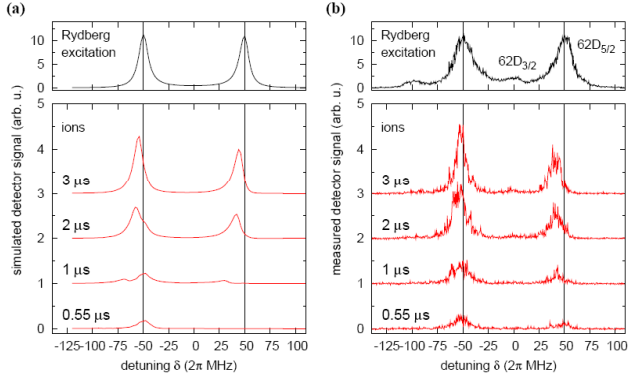}
		}}
\end{figure}

Specific interatomic distance $R$ leads to specific energy position $E$ in
the band structure especially in the  $E=C_6/R^6$ van der Waals
behavior. Therefore, by choosing the detuning of a probe laser, specific
pair distance could be selected. This very attractive possibility can 
be observed using a very simple laser scheme. Indeed, in a two-step
excitation scheme an Autler-Townes energy splitting create a detuning for
the excitation step toward the Rydberg atoms as already shown in figure \ref%
{fig:Rost}  \cite{2007PhRvL..98b3002A}. In other word pair of atoms at
the distance $R = (C_6/E)^{1/6}$, where $E$ is the Autler-Townes energy
shift can be specifically excited.  The anti-blockade has been experimentally observed recently on a
random distributed MOT sample \cite{2010PhRvL.104a3001A} and results are
presented in figure \ref{fig:antiblokade}. Data clearly show that, when the
coupling energy matches the interaction energy of the Rydberg long-range
interactions, the otherwise blocked excitation of close pairs becomes
possible. The experiment, open the way to address specifically distance
between Rydberg atoms.

\subsection{Dipolar blockade}

\subsubsection{Experimental observation}

If the van der Waals blockade, due to $1/R^6$ type of potential interaction, has been observed in 2004,
the dipole blockade itself, i.e. with $1/R^3$  interaction,
has been observed only in 2006 by our group; in
two different experiments, FRET \cite{2006PhRvL..97h3003V} and
permanent dipole interaction \cite{2007PhRvL..99g3002V}, both using external
electric field.

\begin{figure}[h!]
\centering
\resizebox{ 0.9\textwidth}{!}{
		\rotatebox{-90}{
		\includegraphics*[34mm,40mm][88mm,228mm]{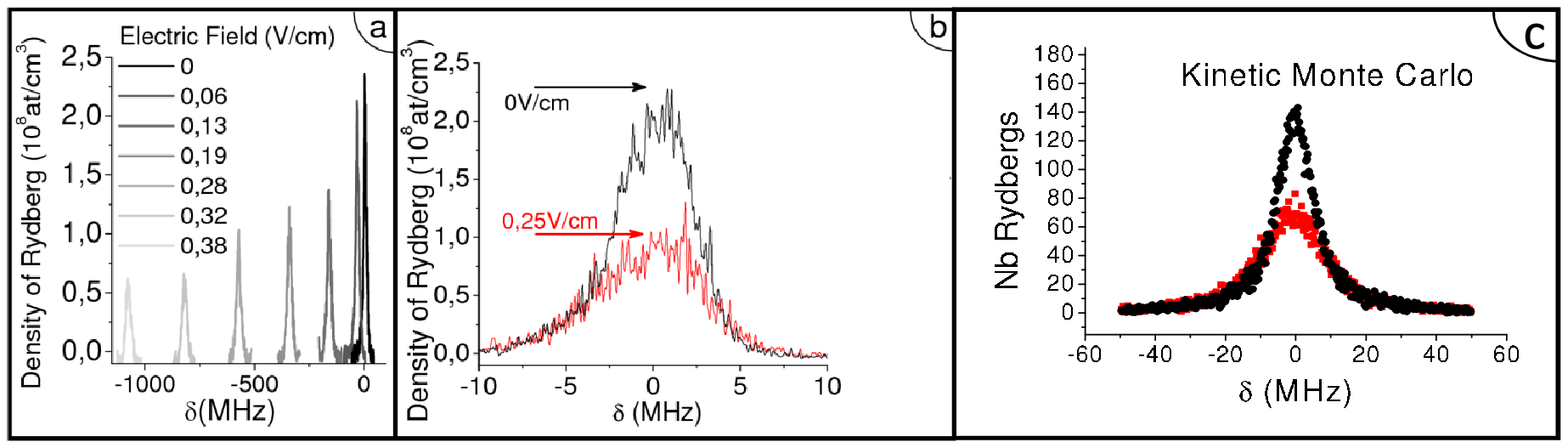}
		}}
\caption{Evidence for the dipole blockade induced by electric field. (a)
Excitation of the spectral lines of the $75p_{3/2}$ level for different
electric fields. The shift of the lines is due to the Stark effect and the
reduced signal is mainly due to the dipole blockade effect. (b) Comparison
of spectral lines of the $70 p_{3/2}$ for two different electric fields. (c)  Kinetic Monte Carlo simulation.
Adapted from \protect\cite{2007PhRvL..99g3002V}.}
\label{Fig:BlocageChamp}
\end{figure}

The evidence for the dipole blockade induced by electric field is shown in
Figure \ref{Fig:BlocageChamp}. The Rydberg atoms are excited from a cloud
produced in a cesium vapor loaded magneto-optical trap. The laser excitation
is provided by three cw resonant lasers simultaneously applied during a time
of $300\,$ns with a repetition rate of $80\,$Hz. The dipole blockade
controlled through Stark effect presents an efficiency of 60\% limited by
the resolution of the excitation ($\sim 5 \,$MHz). In order to avoid any
discussion concerning ionic effects, the experiment has been performed at
relatively low atomic density, using a short time excitation with quite low
laser intensity resulting to less than one ion present per shot.

\subsubsection{Mean field and nearest neighbor approximation.}

As shown in Figure \ref{Fig:BlocageChamp} c) model of the excitation including
the dipole-dipole interaction of each atom with its closest Rydberg neighbor
gives good agreement with the dipole blockade data. The model is the
following. An atom $i$ can be excited in a Rydberg state if the excitation
linewidth (taken into account the laser linewidth, laser broadening and
natural lifetime) $\hbar \Delta_{\mathrm{exc}}$ is smaller than the
dipole-dipole interaction $V = \sum_j V_{ij}$, which is the sum of all
possible interacting with neighbors.  Monte Carlo
simulations can be performed to choose the interatomic distances \cite{Amo08}%
. Assuming rate equation for laser excitation kinetic Monte Carlo method has been chosen because it is very general,
faster than usual Monte Carlo methods, and gives exacts solutions of any
real system evolving through a master equation (or rate equations).

However, simplifications on the $\sum_j V_{ij}$ calculation can be performed
to have an analytical model giving qualitative predictions.

As previously discussed, in the isotropic Van der Waals case $V_{ij} \propto 
\frac{1}{R_{ij}^6}$ and it is possible to replace the sum $\sum_j V_{ij}$ by
an integral which leads to a mean field picture \cite{2004PhRvL..93f3001T}.
This explains why the blockade sphere picture is well adapted to the
isotropic van der Waals isotropic interaction. To be precise, a small angular dependence is still present in the
Van der Waals interaction and the blockade sphere is indeed more a blockade ellipsoid (see Figure 13 of Reference \cite{2008PhRvA..77c2723W}). 
However, blockade sphere
picture and mean field approach has to be used with care for the anisotropic dipole-dipole
interaction. The main reason comes from the fact
that  the angular average of $V_{ij}$ is zero, because $\int_{\theta=0}^{\theta=%
\pi} (1-3 \cos^2 \theta) \sin \theta d\theta =0$.
More precisely, in a mean field coarse grain approach,  the individual atomic position are smooth out and  the  sum  $\sum_{j\neq i}{V_{ij}}$ is replace by an integral and a simple mean field approach would simply gives zero. Consequently, a mean field type of 
treatment  has to be done carefully for instance by separating 
 the nearest
neighboring interactions in the sum.  
Indeed,
a Monte Carlo study \cite{Amo08} clearly indicates that the nearest
neighboring interaction plays an important, if not a dominant, role.
Similarly, F. Robicheaux \cite%
{2008PhRvA..78d0701S} has shown that the pair fluctuation at small separation
is the dominant factor contributing to line broadening. Finally 
in the Förster case the pairs of very close atoms play a particular role for
the observation of broad resonances. More precisely, at exact resonance the interaction has
a finite spatial range when out of resonance the interaction is peaked at
some distance \cite{2009PhRvA..80e2712C}.

Consequently, in a simple estimation of  $\sum_j V_{ij}$ we can isolate the nearest neighbor atom
from all other atoms $j$ having larger internuclear distance. For these far
away atoms the values of $R_{ij}$ are regularly spaced so the angular
average can be performed and leads to a mean field value, which is simply
zero for atoms in the center of a spherical ensemble. For a uniform
distribution of atoms, the distribution of the distance of the nearest
neighbor has Erlang distribution\footnote{%
The probability to find a $k^{th}$ nearest neighbor at a distance $R$ is
given by the Erlang distribution $4\pi R^2 \frac{3}{4\pi k!}\frac{{%
(R^3)^{k-1}}}{(R_0^3)^{k}}e^{-(\frac{R}{R_0})^3}$ and $R_0=(\frac{4\pi n_{%
\mathrm{Ryd}}}{3})^{-\frac{1}{3}}$.}. For simplicity we could assume a single
nearest-neighbor distance $R_0$ given by $R_0^3 n_{\mathrm{Ryd}} \sim 1$. Thus $V
\sim \mu(F)^2/(4 \pi \varepsilon_0 R_0^3) \sim \mu(F)^2 n_{\mathrm{Ryd}%
}(t)/(4 \pi \varepsilon_0 )$ increase with time as the Rydberg density ($n_{\mathrm{Ryd}}$).
Finally the maximum achievable Rydberg density is simply $n_{\mathrm{Ryd, Max%
}}$ given by: $\hbar \Delta_{\mathrm{exc}} \sim \frac{\mu(F)^2 n_{\mathrm{%
Ryd, Max}}}{4 \pi \varepsilon_0 } \sim n^4 \frac{\mu_{pd}^2 F^2}{%
\Delta_{pd}^2/4} \frac{ n_{\mathrm{Ryd, Max}}}{4 \pi \varepsilon_0 } $ This
model is enough to qualitatively explain the results of Figure \ref%
{Fig:BlocageChamp} and especially the fact that  $n_{\mathrm{Ryd, Max}}$ decreases when $F$ increases.

If the angular average of the $1/R^3$ part of the potential is zero, the $%
1/R $ retarded interaction  between the dipoles has
an angular dependence such that the interaction does not vanish when
integrated over the sample. 
Indeed
 when all dipoles are aligned we can
define the angle $\theta $ by $\cos \theta =\left( \vec{n}.%
\vec{\mu }\right) $, and the retardated (generalisation of the hamiltonian given by Eq. \ref{H_12_ret}) becomes:
\begin{equation}
H_{12}=\frac{\mu _{1}\mu _{2}}{4\pi
\varepsilon _{0}}\left\{ -\frac{k ^{2}\cos \left( k R\right) \sin
^{2}\theta }{R}+\left( 1-3\cos ^{2}\theta \right) \left[ \frac{\cos \left(
k R\right) }{R^{3}}+\frac{k \sin \left( k R\right) }{R^{2}}\right]
\right\} ,
\end{equation}%
This formula is  derived using quantum electrodynamics in the appendix in Eq.
(\ref{eq_ham_finale}).

Thus, for the $1/R$ interaction atoms which are farther apart are
relatively more important than they are for the static $1/R^3$ dipole-dipole
interaction. An interesting feature of this interaction, which up to now has
never been studied, is that it is a true collective, many atom interaction
rather than many binary (nearest neighbors) interactions.

\subsubsection{$1/R^3$ behavior.}

\begin{figure}[h!]
\centering
\resizebox{ 0.7\textwidth}{!}{
		\rotatebox{-90}{
		\includegraphics*[34mm,57mm][200mm,260mm]{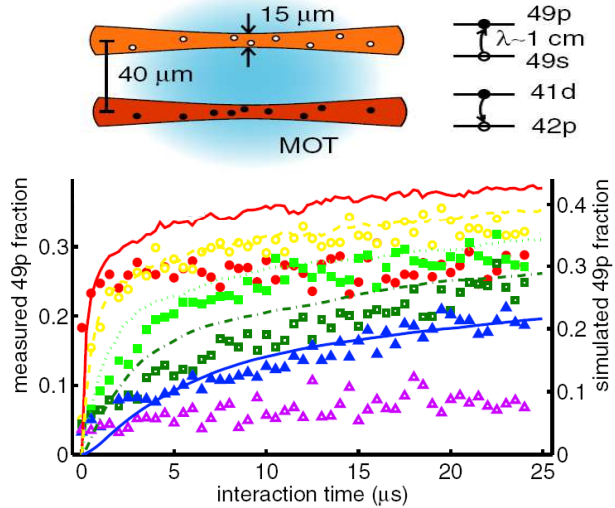}
		}}
\caption{Spatial and time resolution of dipole-dipole interaction between
Rydberg atoms. Upper part, schematic of the Förster ($41d_{3/2} + 49s_{1/2}
\rightarrow 42p_{1/2} + 49p_{3/2}$) experiment. Lower part: measured
production of the $49p$ state as a function of the interaction time for
several separation of the cylinders: from $d=0\,\protect\mu$m (red) to $20$, 
$30$, $40$, $50\,\protect\mu$m and $\infty$ (purple). A full many-body
quantum calculation, performed by F. Robicheaux, (solid line) reproduces the
main features of the experiments, mainly the fact that the transfer rate
scaled as $d^{-5/2}$ with the effective separation distance $d$. Adapted
from \protect\cite{2008PhRvL.100x3201V}.}
\label{fig:Heuvel}
\end{figure}

For non spherical ensemble the angular averaging of the $1/R^3$ interaction is not zero and the results
can be very different. This as been demonstrated by H. B. Van Linden van den Heuvell's  group (Amsterdam) \cite%
{2008PhRvL.100x3201V} where the authors used, not permanent dipole
interaction, but the $41d + 49s \rightarrow 42p + 49p$ Förster resonance.
The interaction in time and space are controlled by varying the laser
excitation beam separation as shown in figure \ref{fig:Heuvel}. The result
clearly show the $1/R^3$ dependence by using two spatially separated atomic
samples. A first volume is for the $41d$ atoms and a second one for the $49s$
atoms. Quantum beat oscillations in $\sin^2(V t) \sim t^2 V^2
\propto t^2/R^6 $, which are expected on the basis of the coherent coupling
between atoms separated by distance $R$ are not observed because they are
average (dephased) when summing over all atoms. However the spatial average $%
\int t^2/ (x^2 +d^2)^3 d x $, along the $x$ coordinate of a laser, leads to a $t^2/d^5$ dependence of the signal
 clearly visible in figure \ref{fig:Heuvel}.

\subsection{Blockade at Förster resonance}

\begin{figure}[h!]
\centering
\resizebox{0.7\textwidth}{!}{
		\rotatebox{-90}{
		\includegraphics*[10mm,2mm][82mm,123mm]{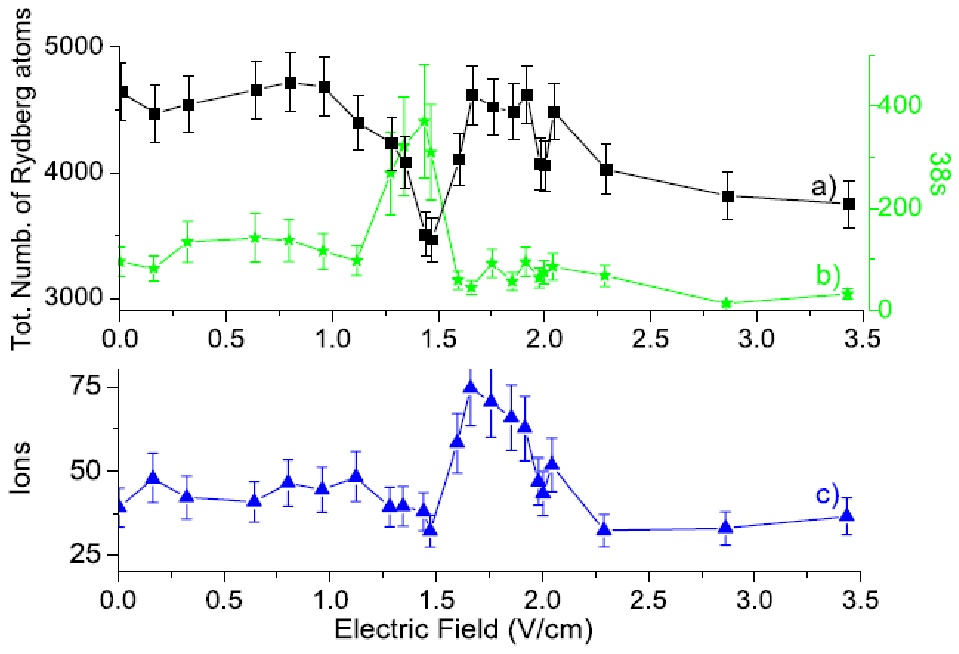}
	}
	}
\caption{Dipole blockade at Förster resonance. The Cs resonance $38p_{3/2} +
38p_{3/2} \rightarrow 38s_{1/2} + 39s_{1/2}$ is study versus the applied
electric field. Upper curve: total number of Rydberg atoms (a, black) and
number of $38s$ atoms (b, green). Lower curve: number of formed ions (c,
blue). Adapted from \protect\cite{2006PhRvL..97h3003V}.}
\label{Fig:BlocageForster}
\end{figure}

The dipole blockade has also been observed at  in a FRET experiment in a 
cesium sample (see figure \ref{Fig:BlocageForster}) \cite%
{2006PhRvL..97h3003V}. The efficiency of the process is characterized by the
minimum of the excitation. It is limited to 30\%, essentially due to the
relatively low $n$ considered ($n < 42$). We notice that the $np$ excitation at
the $np+np\rightarrow ns + (n+1)s$ Förster resonance corresponds to an efficient transfer towards the
levels $ns$ an $(n+1)s$ levels. Few ions are present and are produced after
the laser excitation mostly due to blackbody ionization. It is interesting
to notice that the number of ions is greater for higher electric field than
the field at the Förster resonance. This result is interpreted as a Penning
ionization process because of the attractive force due to dipole-dipole
interaction. A clear evidence of such effect will be exposed in the next
section.

Complementary to this dc electric-field-induced resonant energy transfer,
the ac Stark effect (Radio Frequency of micro-wave) can also be used to
induce degeneracy and resonant energy transfer between cold Rydberg atoms.
This has been clearly demonstrated by H. B. Van Linden van den Heuvell's
team using RF fields in spatially separated volumes \cite%
{2008PhRvA..78f3409T} as well as by J. D. D. Martin's group using a
microwave field \cite{2007PhRvL..98t3005B,2008JPhB...41x5001P}. This method
for enhancing interactions has more flexibility due to the possibility of
varying the applied frequency in addition to the amplitude as in the dc case.

\subsection{Coherence studies of Rydberg sample}

Narrowband excitation opens the way to study in more detail coherent effects
with Rydberg atoms. For instance, adiabatic transfer becomes possible and have indeed been
observed \cite{2004APS..DMP.D1026K,2005PhRvA..72b3405C}. We could also
mention the work based on Coherent Population Trapping or on the
Electromagnetically Induced Transparency (EIT) technique, which have been
first demonstrated by C. Adams's group (Durham, UK) \cite%
{2008JPhB...41t1002W} allowing to have narrow line information \cite%
{2009arXiv0912.4099S} even in a vapor cell room temperature sample \cite%
{2007PhRvL..98k3003M}.

Study of coherence effect inside an ultra-cold Rydberg gas have been
pioneered by T. Gallagher's group which observed a density-dependent
dephasing attributed to the inhomogeneities in the exchange interactions
using Ramsey's type of spectroscopy \cite{2002PhRvA..65f3404A}. In 2005,
using time-resolved narrow-band deexcitation spectroscopy, the Orsay's group
studied similar coherence as well as the destruction of the Rydberg sample
coherence induced by the motion of atoms due to dipolar forces \cite%
{2005PhRvL..95w3002M}.
A clear indication of the modification of the coherence of excitation has
been reported in \cite{2008PhRvL.100k3003J}. With one atom in the trap the
authors observed Rabi oscillations but, when more than one atom is in the
trap they observed loss of contrast of the Rabi oscillations due to the van
der Waals interaction of the excited atoms. These experiments could not
distinguish between real decoherence or simple dephasing. Spin-echo, or
phase reversal technique, first use by T. Pfau's team \cite%
{2007PhRvL..99p3601H,2008PhRvL.100a3002R,2009NJPh...11e5014R}, then
experimentally and theoretically study by other teams \cite%
{2008JPhB...41s5301H,2009NJPh...11d3006Y,2010PhRvA..81b3406W}, have revealed
that in a frozen gas, the evolution can indeed be hamiltonian and the coherence 
still present even if hidden for standard detection methods. Finally, double
crossing Landau-Zener-Stueckelberg oscillations in the dipole-dipole
interaction between Rydberg atoms has been observed recently (single
crossing have been studied by Pillet's group \cite{fioretti1999a,vanhaecke2010}), in H. B.
van Linden van den Heuvell's group, using an externally applied
radio-frequency field, proving coherent dipole-dipole interaction during at
least $0.6\,\mu$s \cite{2009PhRvA..80f3407V}. 

Conclusions of all these
studies are that the dephasing rate is found to increase with density and
that for a typical Rydberg density of $10^9\,$at/cm$^3$ decoherence occurs
in less than $1\,\mu$s. This obviously put limits concerning the capability
to realized quantum gate in many atoms sample \cite{2009arXiv0909.4777S}.

Another collective or cooperative effect of cold Rydberg sample, similar to
the behavior of Frenkel excitons \cite{PhysRevB.51.7655}, concerns their
superradiant behavior as observed by many groups \cite%
{2007PhRvA..75c3802W,2008PhRvA..77e2712D}. As the theory
predicts \cite{2005quant.ph..9184Y,2008PhRvA..77c3844B} small atom systems
show the increased emission generated by the strong dipole-dipole
interaction. However, for large numbers of atoms the effect of the
dipole-dipole interaction on collective emission is reduced. These non
linear effects are closely related to strong coupling to light observed in
micro-wave of black-body coupling in Serge Haroche's group \cite%
{2005PhRvL..94k3601M,2008EPJST.159...19H}. Other non linear effect such as the already mentioned
EIT or four-wave mixing have been observed and are promising for studying
quantum effects in blockaded atom cloud \cite{2008PhRvA..78f3830B}.

\subsection{Quantum equilibrium behavior of a Rydberg sample}

\begin{figure}[h!]
\centering
\resizebox{0.7\textwidth}{!}{
		\rotatebox{-90}{
		\includegraphics*[15mm,81mm][191mm,223mm]{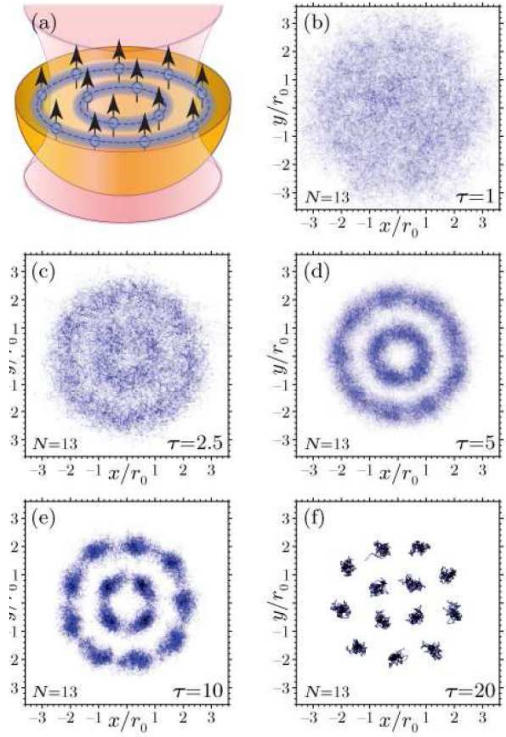}
		}}
\caption{Ground state estimates of many Rydberg atoms system. In a possible
experimental setup (scheme in a): Rydberg-dressed atoms are confined to 2D
by a strong confining laser beam, with dipoles polarized perpendicular to
the plane. In-plane harmonic confinement is provided e.g. by the beam waist.
(b-f) Monte Carlo snapshots of the density of particles for $N = 13$
dipoles, for several strength $\protect\tau$ of the dipole-dipole
interactions. (b) superfluid; (c) supersolid; (d-e) ring-like crystals; (f)
classical crystal. From \protect\cite{2009arXiv0904.2735P} see also \cite{Pupillo:arXiv1001.0519}.}
\label{Fig:dipPhaseZoller}
\end{figure}

What are the equilibrium quantum properties of a dipolar gas sample? Since
2000 several theoretical studies \cite{2000PhRvL..85.1791S}, especially by
G. Shlyapnikov, P. Zoller, M. Lewenstein or L. Santos' groups, have tackled
this problem of dipolar gases near quantum degeneracy \cite%
{2002PhST..102...74B,revue_dip_pfau}. The phase diagram of such extremely
cold systems is complex and depends on the ratio of dipole-dipole
interactions over contact (scattering) interactions. Such studies have
obviously a lot of connection with solid state physical systems \cite%
{2006PhRvL..96s0404S,2006PhRvL..96h0405K}. Similar dipole-dipole studies can be performed using cold
polar molecules \cite%
{krems2008,Bell2009,Review_Ye2009,2009RPPh...72h6401D,2009arXiv0904.2735P}
or magnetic dipolar gases, such as beautifully observed in strongly
perturbed anisotropic expansion of a Chromium condensate by Tilman Pfau's
group \cite{2005PhRvL..95o0406S,2008NatPh...4..218K}. Rydberg gases can be
seen as frontier system with huge dipole-dipole interaction.
"Thermodynamical" equilibrium is rich (see Figure \ref{Fig:dipPhaseZoller})
and has strong similarity with spin magnetism because the ground state of
the Hamiltonian describing the system exhibits a similar phase transition in
both systems \cite{2008PhRvL.101y0601W,2009arXiv0904.2735P,Pupillo:arXiv1001.0519}. However, the
reduced lifetime and inelastic collisions have prevented, up to now, to
study such equilibrium type of behavior. For instance, Tilman Pfau's group,
study the Rydberg interaction inside a Bose-Einstein condensate \cite%
{2008PhRvL.100c3601H} where only the high density play a role but no special
quantum effect due to the coherence properties of the Condensate have been
observed.

An interesting idea, which can help to study equilibrium states, is to combine the large dipole
given by Rydberg states with the long lifetime given by the ground state
atoms \cite{2000PhRvL..85.1791S}. It consist to induce a the
large dipole moment to long lived ground state atoms by coupling them weakly
(i.e. by dressing them out of resonance) to Rydberg
states \cite{2009arXiv0904.2735P}. However controlling the amount of Rydberg
population necessary seems challenging \cite%
{2007PhRvA..75a2304W,2009arXiv0909.4777S}.

\subsection{Multi-body effects in the modelization of an ultracold Rydberg
gas}

Before describing  difficulties presented by the many-atomic treatment,
we first  present what is the convenient hamiltonian description of the physics.

\subsubsection{Laser Rydberg excitation of a large ensemble of cold atoms}

The laser excitation of several atoms is much more complex to treat that the evolution of two atoms (which is derived in detail in the appendix \ref{appendix}), even if we stay in a frozen gas approach. 
For instance in the  example of the $np+np \leftrightarrow ns + (n+1)s$ reaction   cesium atoms, the atoms can be excited in states
$\left\{
\left\vert r^{(0)} \right\rangle ,\left\vert r^{(1)} \right\rangle ,\left\vert
r^{(2)}\right\rangle \right\} = \left\{
\left\vert n s \right\rangle ,\left\vert n p \right\rangle ,\left\vert
(n+1) s\right\rangle \right\}$.  We also the Förster hamiltonians terms corresponding to an exchange of internal energy, $ np + np \leftrightarrow ns + (n+1)s  $ and $ np + np \leftrightarrow (n+1) s + n s  $ between two atoms $i$ and $j$ ($i$ being the first one and $j$ the second in the equations).
We have also migration hamiltonian terms,
corresponding to an exchange of excitation, $ i(ns) + j(np)  \leftrightarrow i(np) + j(ns) $, $ (n+1)s + np \leftrightarrow (n+1)s + np$  and $ (n+1)s + ns \leftrightarrow (n+1)s + ns$.
Thus the total hamiltonian is very rich, but
all reactions between an atom $i$ and an atom $j$ 
 can be treated. 
 
 In general, if all the possible Rydberg states in the $i=1, \cdots, N$ atoms system are labeled by $r^{(n)}$,  
 the
 hamiltonian (containing even off resonant terms) is given by the generalization of Eq. \ref{H_1_and_2}
\[ H= \sum_{i}  \sum_{n } \hbar \omega_{r ^{n}} |r_i^{(n)} \rangle \langle r_i^{(n)} | + \sum_{ i,j}  \sum_{n,n';n'',n'''} 
H_{12}^{r_i^{(n)} r_i^{(n')}; r_i^{(n'')} r_i^{(n''')}}\]

\subsubsection{Optical Bloch equations to study laser Rydberg excitation}

If we want to include the radiative lifetime it is necessary to describe the system using the  time evolution of the density matrix $\rho$ given by the
optical Bloch equations. Detail treatment has been given in \cite{Amo08,2007PhRvA..76a3413A} and we are not going to enter in any detail here,  but the next sections will highlight some of the difficulties.
Very briefly, taking the trace over all the atoms
except the one labeled $i$ in the optical Bloch equations, gives the
evolution of the density matrix $\rho_i$ for the particle $i$. The interaction term
or shift in energy for the atom $i$ due to the interaction with its
neighboring atoms is $\sum_{j\neq i}{Tr_{j}[H_{ij},\rho_{i,j}]}$, with $%
H_{ij}$ the dipole-dipole interaction and $\rho_{i,j}$ the two-body density
matrix for atoms $i$ and $j$. 
As correlations appear during the excitation, the state of the system does
not remain a product state. However if the probability of excitation of a
ground state atom into a Rydberg state is small, on the order of a few percent,
and as long as the product of the individual density matrices is small, we
can use the Hartree-Fock approximation where $\rho_{i,j} \sim \rho_i \otimes \rho_j$.
Thus the interaction term can be written, schematically, as 
\begin{equation}
\sum_{j\neq i}{Tr_{j}[H_{ij},\rho_{i,j}]}=(\sum_{j \neq
i}V_{ij}\rho_{j_{r r}})(\rho_{i_{g r}}|g \rangle_i {}_i\langle
r|-\rho_{i_{eg}}|r \rangle_i {}_i\langle r|)  \label{eq:interaction}
\end{equation}
which is simply a shift of the Rydberg level for the atom $i$. The
population in the excited state $\rho_{j_{r r}}$ is can then be replaced by a local
mean value $\rho_{r r}(\vec{R})$ considered at different positions $\vec{R}$
over the whole atomic cloud. As we have seen, a naive (mean field)
estimation for $\rho_{j_{r r}}$ could lead to wrong estimations and it is
often better is to consider separately the nearest neighbor Rydberg atom
from the other atoms. The shift in energy relies on the local density
(gaussian distributed) $\rho_0(\vec{R})$ of the atoms in ground state. The
shift for the atom $i$ is given by the interaction term of Eq. (\ref%
{eq:interaction}) which is finally proportional to $\rho_{r r}(%
\vec{r})\rho_0(\vec{r}) $. We can then use this approximation to
solve equation the equation for an atom $i$.

\subsubsection{Full quantum mechanical treatment vs two-atom or mean field approach}

We have previously only briefly mentioned some basic results concerning the
modelization of an ultracold Rydberg gas, but we would like to summarize
here some of the approach or problems that theory has to face. Indeed,
theoretical study of an ultracold Rydberg gas is very complex due to the
numerous effects presents such as black-body radiation, quantum collective
effects, dipolar forces, excitation transfer or dipole orientation effects to list few of them.
All this explains why almost all experiments have tried to interpret their
data using simple two-body, or one body plus mean field, picture. First,
Akulin's study \cite{1999PhyD..131..125A} as well as Frasier's one 
\cite{1999PhRvA..59.4358F} have considered subsequent two body interactions
to interpret data concerning diffusion of Rydberg excitation. The
Connecticut group was the first one to proposed a many body (mean field)
approach to study the Van der Waals blockade \cite{2004PhRvL..93f3001T}.
This approach has been modified to take into account the nearest neighbor
dominant effect in dipole blockade case by the Orsay group \cite%
{2007PhRvL..99g3002V,Amo08}. The same group has demonstrated, in a similar
way to what has been performed by Jan Michael Rost's group (Dresden, Germany) 
\cite{2007PhRvA..76a3413A}, that this simple approach can be deduced from a
more complex analysis of the density matrix evolution has described previously.

\begin{figure}[h!]
\centering
\resizebox{0.7\textwidth}{!}{
		\rotatebox{-90}{
		\includegraphics*[30mm,77mm][171mm,190mm]{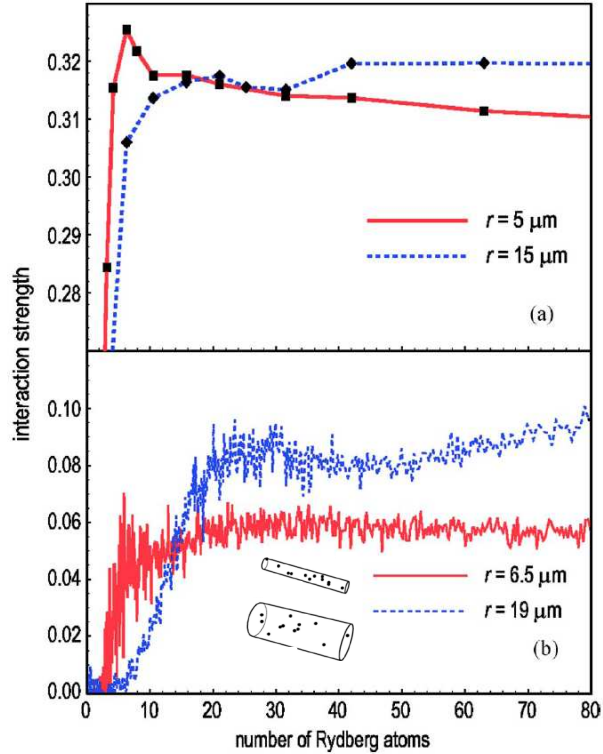}
		}
		}
\caption{Evidence of many body effects and their dependence with
dimensionality. a) results of the simulations of the $31d + 31d \rightarrow
33p + n= 29 $ transfer. The interaction strength (fraction of atoms that
interacts) behavior reproduces qualitatively the experimental features of
the data shown in part b) taken at two different excited volumes. Even though
the density is higher in the smaller, one dimensional volume, the
interaction is stronger in the sparser but more three-dimensional volumes.
Adapted from \protect\cite{2006PhRvA..73c2725C}.}
\label{fig:manybodyNoel}
\end{figure}

Francis Robicheaux studies carefully quantum effects in function of the atom
number in the sample \cite%
{2004PhRvA..70d2703R,PhysRevA.72.063403,2008JPhB...41d5301H}. Similar study
have been performed (for $N<10$) by Matthias Weidemüller's group \cite%
{2006EPJD...40...37W}. 
Especially in the Förster resonance case, non perturbative
many-atom Hamiltonian agree more with experimental results than sums over
pairwise atomic potentials \cite{2009PhRvA..79d3420Y}. 

The fact that, the migration of
the products of the reaction by exchange of Rydberg excitation between
atoms, has to be taken into account or not is still a matter of debate \cite%
{review_gall_pillet}. Indeed, the interpretation of the line broadening mechanism
has  been recently reconsidered by F. Robicheaux \cite%
{2008PhRvA..78d0701S} and, contradictory to the interpretation usually based on the diffusion of
excitations \cite{2004PhRvA..70c1401M}, by turning off the diffusion process in the theoretical model no important
change in the result was visible in the spectral linewidth.

Finally, the many body effects
can also depend on the dimentionality of the sample as shown in the figure %
\ref{fig:manybodyNoel}. At small radius, where the sample is nearly
one-dimensional, many-body interactions are suppressed. But, when the sample
becomes three-dimensional (large radius) many-body effects are apparent \cite%
{2006PhRvA..73c2725C}. Furthermore, in a sample with larger number of atoms,
the transfer occurs faster despites a lower density \cite%
{2004PhRvL..93o3001C,2006PhRvA..73c2725C}.

All this, clearly point out that, contrary to the van der Waals or the
static dipolar interaction, where simple binary picture are usually enough to
obtain precise results, in the Förster case it is necessary to include
multiple atoms and to solve the full many-body wave function \cite%
{2005JPhB...38S.309W,2008PhRvA..77c2723W}. If the full many-body wave
function is necessary to accurately model Förster experiment, the number of
atoms to be included in the simulation to accurately model the experiment is
still unclear (it varies from 4 to 12 \cite%
{2007NuPhA.790..728R,2009PhRvA..79d3420Y,2009PhRvA..80e2712C}). However,
only two of the states seems to play a dominant role suggesting that some
simplification of the analysis may be possible by considering the atoms
collectively \cite{2009PhRvA..80e2712C}.

\section{Control of Penning ionization}

\label{sec_dyn}

If Penning ionizations play a key role in plasma formation they perturb the
observation of blockade excitation, because the electric field created by
one ion is much bigger than the one created by a Rydberg dipole. It is
indeed very difficult to distinguish Rydberg blockade effects from ion plasma dynamics 
\cite{2004PhRvL..93p3001S,2006EPJD...40...27L,2006EPJD...40...37W,Amo08}. This has been study
in our group in Orsay,
by
adding a simple N-body integrator (Verlet) to the Kinetic Monte Carlo
simulation. We were able to describe dynamic processes in space
and time and clarifies the role of collisions leading to ions in some
"blockade" experiments \cite{Amo08}.

\subsection{Evolution toward a plasma}

A cold Rydberg gas can spontaneously evolve into a plasma \cite{vitrant1982}. The process is the following.
Even a very slow ionization process, produces cold ions.
At some point their macroscopic space charge traps all subsequently
produced electrons \cite%
{ref2000PhRvL854466R,killian2003,ref2007PhR...449...77K}. For an electron
temperature $T_e$, the number of ions $N_{\mathrm{ion}}$ needed is simply
given by the relation $\frac{N_{\mathrm{ion}} q_e^2}{4 \pi \varepsilon_0 l}
\sim k_B T_e $ in a sample of size $l$. A trapped electron can then collide
with a Rydberg atom present and ionize it leading to a free electron with a
kinetic energy on the order of twice the initial Rydberg binding energy \cite%
{2005JPhB...38S.333R}. This leads to a collisional avalanche which rapidly
redistributes the population initially put into a single Rydberg state. At
the end most of the Rydberg atoms are ionized. If Rydberg atoms can be
ionized, ultracold plasma also formed Rydberg atoms in a back and forth
evolution \cite{killian2001,2003JOSAB..20.1091G}. Surprisingly, the Rydberg
atoms in the plasma has a profound analogy with binary stars in star cluster 
\cite{2005MNRAS.361.1227C}. For instance with the same Heggie's law: soft
binaries get softer and hard binaries get harder. Indeed, if the Rydberg
(binary) binding energy is roughly higher than a free electron (star)
kinetic energy then the Rydberg atoms (binary stars) are driven to lower
states.

In an initial pure Rydberg sample, the origin of the initial ions can be due
to different processes such as for instance collision due to hot background
atoms or ionization trough blackbody radiation. However, if exists, the
dominant mechanism for this ionization is that pairs of atoms excited to
attractive diatomic potential curves collide resulting in the ionization of
one of the atoms, the second atom being driven to a lower state (Penning
ionization process) \cite{walz2004,Li1}.

\subsection{Penning ionization between Rydberg atoms}

\begin{figure}[h!]
\centering
\resizebox{0.7\textwidth}{!}{
		\rotatebox{-90}{
		\includegraphics*[56mm,37mm][126mm,202mm]{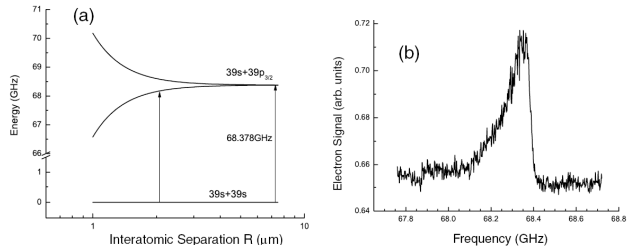}
		}}
\caption{Penning ionization due to attractive potential curves. (a) Energy
levels for the Rb pair states 39s39s and 39s39p as a function of
internuclear separation (b) Plasma electron signal observed as a function of
microwave frequency driving the $39s39s \rightarrow 39s39p$ transition; only transitions to the attractive curve of (a) are
observed. From ref. \protect\cite{2005PhRvL..94q3001L}.}
\label{fig:Penning_Gall}
\end{figure}

The key role play by the dipolar forces in the ionization process have
clearly been observed in T. Gallagher's group who populated Rb$(ns)$ states
with a pulsed blue laser and then drive microwave transitions from the
molecular $ns ns$ state to the $ns np$ state \cite{2005PhRvL..94q3001L}. The authors monitored the ion signal during the
microwave frequency was scanned, results are
shown in Figure \ref{fig:Penning_Gall}. Microwave excitation to the attractive
curve results in ionization, but when it is to the repulsive curve it does
not.

If the atoms are drawn together along an attractive curve and ionize,
dipole-dipole potential energy is converted into kinetic energy of the
resulting atom. This effect has been observed in reference \cite{2006PhRvA..73b0704K}. Assuming an $%
C_n/R^n$ interaction potential curve, the collisional time of two atoms
(with a reduced mass equal half the individual mass $M$) initially separated
by a distance $R_0$ can be very simply estimated to be $T = \int_0^{R_0} 
\frac{1}{\sqrt{\left( \frac{C_n}{R^n} -\frac{C_n}{R_0^n} \right) 4/M } } d R$%
. For resonant dipole interaction $C_3 \sim n^2 a_0$ and this becomes $T
\sim 20 \mu$s$\times \frac{\sqrt{M(amu)R_0^5(\mu m)}}{n^2}$. For $n = 50$
rubidium atoms and an initial separation of $5\,\mu$m the time is $4 \mu$s 
\cite{2005JPhB...38S.333R}. This straightforward calculation of the time
required for the atoms to move along the attractive path and to ionize
agrees well with experiments.

\begin{figure}[h!]
\centering
\resizebox{0.7\textwidth}{!}{
	\rotatebox{-90}{
		\includegraphics*[100mm,79mm][188mm,211mm]{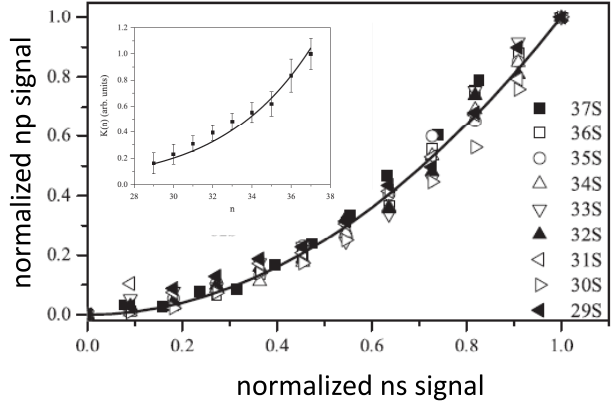}
	}
	}
\caption{Observation of dynamical Stark effect during the laser excitation.
Quadratic (solid line) dependence of $np_{3/2}$ Rydberg signal after
excitation of cold rubidium $ns$ Rydberg atoms for several $n$ values.
Because the interaction between $ns$ atoms is repulsive, the presence of $%
np_{3/2}$ atoms are not coming from binary collision but comes from direct
laser excitation facilitated by the ac Stark shift created during the laser
excitation and by the dipole-dipole mixing between the $ns+ns$ and the $%
np+np $ levels. This hypothesis is attested by the inset showing the
expected (for the theory see \protect\cite{2009PhRvL.102u3201N}) $K(n)
\propto n^{8.5}$ dependence of the rate $N_{np} = K(n) N_{ns}^2$. Adapted
from ref. \protect\cite{2009PhRvL.102u3201N}.}
\label{fig_twoPhoton}
\end{figure}

However, latter studies have raised unexpected results. Even in presence of
repulsive forces, as in the van der Waals Rb$(n s)$ case, Penning ionization
still occur even if slower than in the case of an attractive interaction ($n
d $ states) \cite{2007PhRvL..98b3004A}. The explanation of this phenomenum
is still matter of debate. Effect of Blackbody radiation transferring population towards states
having attractive forces have been suggested \cite%
{2007PhRvA..76e4702A,2008PhRvL.100l3007R,2009EPJD...53..329A}, but the
predicted rate seems too low to ensure sufficient ionization \cite%
{2008PhRvA..78d0704V}. Irregularity in the arrangement of the atoms can also
play a role by leading to an acceleration of the dynamics \cite%
{2008NJPh...10d5030A}. In recent experiments, a nonlinear dependence on the
laser power has been observed (see figure \ref{fig_twoPhoton}) suggesting
that more than one atom is excited simultaneously and that the role of
dynamical Stark effect during the laser excitation must also be considered 
\cite{2009PhRvL.102u3201N}. Similarly, to interpret the fast ($100\,$ns)
ionisation of Rb$(n=88)$ at a density of $5 \times 10^{10}\,$cm$^{-3}$
observed in reference \cite{2008PhRvL.100d3002T}, multiple atom absorptions
has been suggested to lead to rapid ionization in a sequence of near
resonant dipole-dipole transitions \cite{2008PhRvL.100d3002T,hahn2000}.

\begin{figure}[h!]
\centering
\resizebox{0.7\textwidth}{!}{
		\rotatebox{-90}{
		\includegraphics*[50mm,43mm][162mm,207mm]{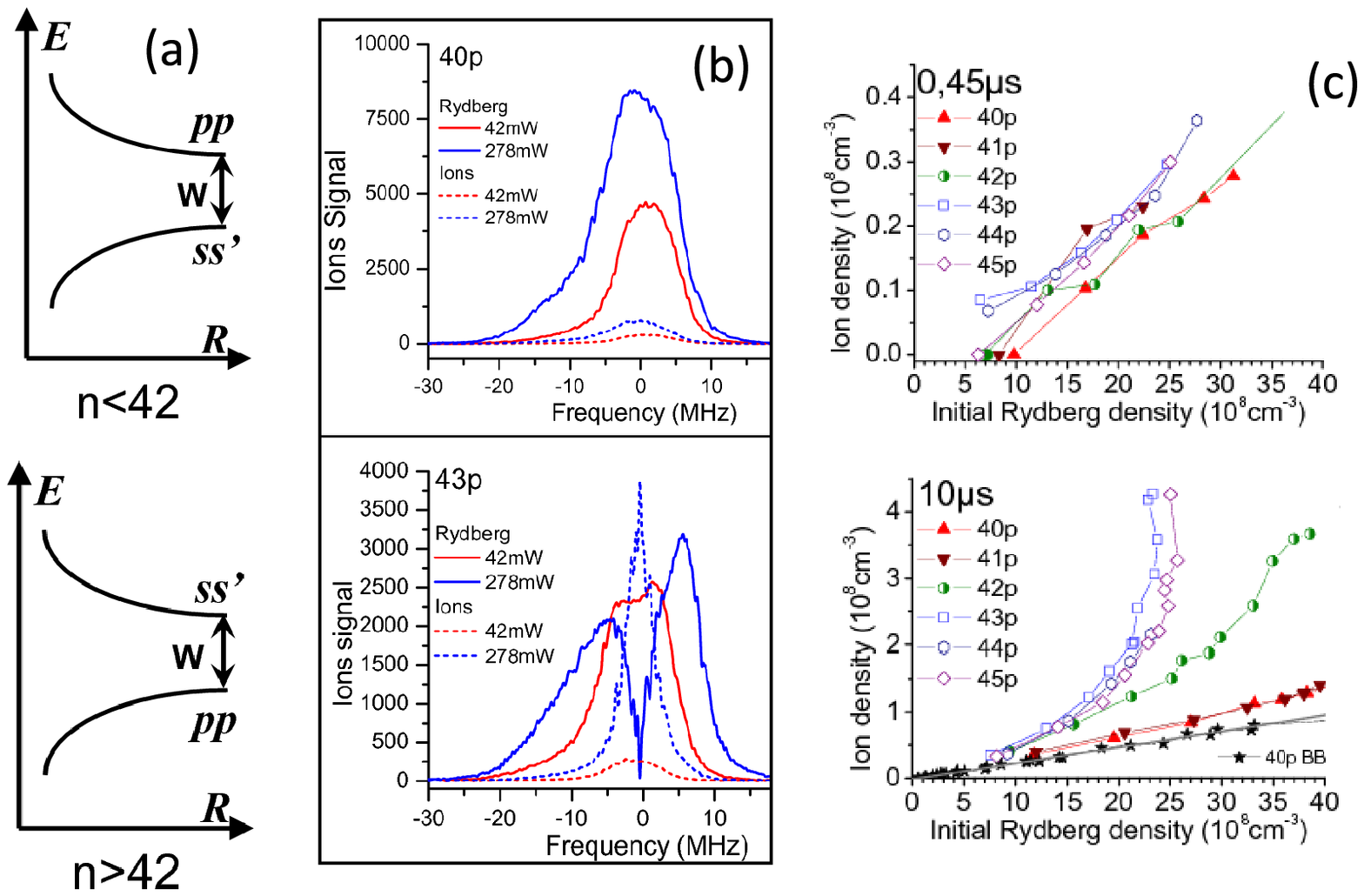}
	}
	}
\caption{Penning ionization on attractive or repulsive potential curves. (a)
Schematic potential curves of a pair of cesium atoms representing, as a
function of internuclear distance, the energy difference (at zero electric
field) between a pair of atoms in state $np,np (pp)$ and $ns,(n+1)s
(ss^{\prime })$. (b) Rydberg and Ion signal with two different laser
intensity ($42\,$mW and $278\,$mW) for (upper part) $40p$ state (repulsive)
and (lower part) $43p$ (attractive). Data are taken after $10\,\protect\mu$s
of free evolution. (c) Ion density as function of initial Rydberg density
for different $n$ states and two delay time (upper part) $0.45\,\protect\mu$%
s and (lower part) $10\,\protect\mu$s the "40p BB" is taken with no $6p$
atoms present. Adapted from ref. \protect\cite{VIT08}.}
\label{fig:PenningCs}
\end{figure}

In an effort to isolate the effect of attractive and repulsive potentials
from other effects, we have examined the ionization of cold Cs$(np)$ atoms
excited with narrow-bandwidth excitation \cite{2008PhRvA..78d0704V}. The interesting
features of the cesium $np$ states are shown in figure \ref{fig:PenningCs}
(a). For $n > 42$ the $np+np$ state lies below the $ns + (n+1)s$ one, while
for $n < 42$ the reverse is true. 
The experimental ionization results, shown in figure \ref{fig:PenningCs}
(b-c), display unambiguous difference between repulsive and
attractive  potentials. For  $n > 42$  (attractive potentials) we observe that the ion production
 become very non linear leading to a the complete ionization of the
Rydberg sample. This is the signature of the formation of an ultracold
plasma. For $n < 42$ (repulsive potentials), the ionization is mainly due
to blackbody radiation but we also suggest possible effect of ion-Rybderg interactions.

\subsection{Exotic molecules in Rydberg systems}

We could not conclude the review of the Rydberg blockade mechanism, i.e. on
the interaction between two Rybderg atoms without mentioning that in
addition to their long range part, in $C_n/R^n$, molecular potential energy
surfaces reveals exotic Rydberg molecules. Indeed, the presence of many
nearly degenerate pairs of states implies that the potential curves exhibit
avoided crossings as a function of the interatomic distance $R$, which
introduces the possibility of molecular states which can be bound. In 2000
C. Greene and H. Sadeghpour's groups predicted very peculiar levels some of them having
permanent dipole moment even without any external field present. They called
such states "trilobite" or "butterfly" state because of the spacial electron
density distribution which has analogous shape compare to trilobite or to
butterfly animals \cite%
{2000PhRvL..85.2458G,2001PhRvA..64d2508G,2002JPhB...35L.199H}. Such
molecular states involved one Rydberg electron surrounding one neutral atom.
The first attempt to observe such states have been performed at Stoors (Connecticut), but
only molecular potential curve crossing have been observed \cite%
{2003PhRvL..91r3002F,2003PhRvL..90n3002D}. The first observation of such
states have in fact been performed in a room temperature experiment \cite%
{2006PhRvL..97w3002G,2009PhRvA..80e2506V} before being clearly observed in
an ultracold ensemble \cite{2009Natur.458.1005B} where spectra of the
vibrational states has even been recorded as well as triatomic molecule or
quantum reflection bound excited dimer states \cite{2009arXiv0912.4058B}.

Interaction between two Rydberg atoms, has been studied in 2002 by Robin Côté%
's team. They predicted pure long range molecules which they call
macro-Rydberg \cite{2002PhRvL..88m3004B}. Similar study adding sometimes
electric, magnetic or many (eventually orderered) atoms interactions have
confirms the existence of many exotic Rydberg molecules \cite%
{2002JPhB...35L.193C,2002PhRvA..66d2709K,2006EPJD...40...65L,2007PhRvL..99k3004M,2006PhRvA..74b0701S,2007PhRvL..98b3002A}%
. Poly-atomic molecules have also been predicted, even in repulsive
di-atomic case because a third atoms can stabilize the system \cite%
{2009PhRvL.102q3001L}. Using a resolution approaching $1\,$MHz Shaffer's
team recently report \cite{2009NatPh...5..581O} the observation of cold
cesium Rydberg-atom macro-molecules bound at internuclear separations of $R
\sim 3-9 \,\mu$m. The molecules are observed using the Coulomb repulsion of
the ions after pulsed field ionization \cite{2007PhRvA..76a1403O}.

\section{Conclusion}

\label{sec_con}

The Rydberg atoms with their huge electron dipole moment offer really a powerful tools for manipulating the interaction between atoms in presence of fields.
Dipole-dipole interaction is the dominant feature of the cold Rydberg gas
and the control of the dipole-dipole forces between Rydberg atoms opens
interesting prospects \cite{review_gall_pillet}). 
The Rydberg dipole blockade offers an efficient quantum engineering for the
entanglement of pairs of atoms and for the realization of quantum gates, quantum simulator or quantum repeater \cite{2009arXiv0909.4777S2010,NatPh...6..382W}.
More and more complex process can be imagined for controlling the mesoscopic system.
Conditional transfer of ensemble atoms will probably be achieved in a near futur,
for instance between two logical states by combining EIT and blockade \cite%
{2009PhRvL.102q0502M}.

The role of the dipole-dipole coupling at the Förster resonances, where
multiple atoms and evolution of the full many-body wave function has to be
solved, opens the way for simulating quantum many body physics  \cite{2007AdPhy..56..243L,2009Sci...326..108B,2010NatPh...6..382W}.
Similarly,
creation of complex many-particle states with atoms or ions in new
geometries, eventually dynamical ones \cite{2008PhRvA..77a3822S}, such as
lattice, make them potentially attractive for the study collective
excitations \cite%
{2008NJPh...10i3009M,2009PhRvA..79d3419O,2009PhRvL.103r5302O}, quantum
random walk \cite{côté2006quantum} or strongly correlated electronic system 
\cite{2008PhRvA..78d3415V}. The fact that the energy levels mainly depend on
the number of excited Rydberg states could be used to experimentally
control, through the laser frequency, the number of excited states or even
to control the interatomic distance to realize a well-defined regularly
spaced Rydberg atomic sample \cite{2007JPhB...40.3693W,2009arXiv0911.1427P}. The use of  the strong
Rydberg blockade regime, or  its breakdown for specific internuclear distances, \cite{2010PhRvL.104a3001A,2009PhRvA..80e3413Q} open
the possibility to excite only  nearest neighboring atoms at well defined distance \cite%
{2010PhRvA..81b3406W}.
This could be combined with the trapping of Rydberg atoms as realized by F.
Merkt's group \cite{2008PhRvL.100d3001H}, or even coupled with laser
ionization to  prepare correlated plasmas \cite%
{2004PhRvL..92o5003P,2004JPhB...37L.183P}. 

This review has only cover the case of alkali atoms. But very attractive fundamental  as well as applicative research can be imagined using, for instance, two-valence electron atoms, such as helium, alkaline earth metals or  ytterbium atoms. For instance, we might imagine using the first electron to excite the atom in a Rydberg state while manipulating the atom using the second valence electron  to image, to laser cool  and even to trap the Rydberg atoms.

\section{Acknowledgments}

This work is performed in the frame of the Institut Francilien pour la
Recherche sur les Atomes Froids (IFRAF), of the Réseau Thématique de
Recherche Avancée "Triangle de la Physique".

\section{Appendix}

\label{appendix}

We derive here the master equation for two atoms excited in Rydberg state, taken into account retardation effect, Lamb shift effect and also (cooperative) spontaneous emission or superradiance. Other treatments can also be found in references
\cite{2005quant.ph..9184Y,2008PhRvA..77c3844B,2010PhRvA..81e3821S}.

\subsection{Dipole-dipole interaction between two two-level atoms}

We consider first two identical two-level $\left\{ \left\vert 1\right\rangle
,\left\vert 2\right\rangle \right\} $ atoms $A_{i}$ $i=\left\{ 1,2\right\} $%
, interacting with the vacuum radiation $V$, in the frame of an electric
dipole transition. The Hamiltonian, $H$, of the whole system $\left\{
A_{i}\right\} +V$ is%
\begin{eqnarray}
H &=&H_{V}+H_{A}+H_{AV} \\
H_{A} &=&\underset{i=1}{\overset{2}{\sum }}H_{Ai} \\
H_{AV} &=&\underset{i=1}{\overset{2}{\sum }}H_{AiV}
\end{eqnarray}%
where $H_{Ai}$ is the hamiltonian of the atom $i$ alone, and $H_{V}$ the
hamiltonian of the vacuum radiation. We have%
\begin{eqnarray*}
H_{Ai} &=&\hbar \omega _{1}\left\vert 1\right\rangle _{ii}\left\langle
1\right\vert +\hbar \omega _{2}\left\vert 2\right\rangle _{ii}\left\langle
2\right\vert  \\
H_{V} &=&\underset{l}{\dsum }\hbar ck \hat{a}_{l}^{+} \hat{a}_{l}
\end{eqnarray*}%
$l=\left\{ \vec{k},\vec{e}_{l}\right\} $ corresponds
to the modes of the vacuum radiation describes by the creation and
anhilation operators $\hat{a}_{l}^{+}$\ and $\hat{a}_{l}$.$\ H_{AiV}$ the
interaction hamiltonian coupling each atom, $A_{i}$, and the vacuum
radiation $V$%
\begin{equation}
H_{AiV}=-\sum\limits_{l}\varepsilon _{l}\left\langle 2\right\vert \vec{d}.%
\vec{e}_{l}\left\vert 1\right\rangle \left[ \hat{a}_{l}\exp \left( i\vec{k}.%
\hat{\vec{R}}\right) -\hat{a}_{l}^{+}\exp \left( -i\vec{k}.\hat{\vec{R}}%
\right) \right] \left\vert 2\right\rangle _{ii}\left\langle 1\right\vert
+h.c.
\end{equation}%
In the frame of the rotating wave approximation, we consider only the part
of the hamiltonian%
\begin{equation}
H_{AiV}^{RW}=-\sum\limits_{l}\varepsilon _{l}\left\langle 2\right\vert \vec{d%
}.\vec{e}_{l}\left\vert 1\right\rangle \hat{a}_{l}\exp \left( i\vec{k}.\hat{%
\vec{R}}\right) \left\vert 2\right\rangle _{ii}\left\langle 1\right\vert
+h.c.
\end{equation}%
with $\varepsilon _{l}=i\sqrt{\frac{\hbar \omega _{l}}{2\varepsilon _{0}L^{3}%
}}$. The non resonant part of the hamiltonian is%
\begin{equation*}
H_{AiV}^{NRW}=\sum\limits_{l}\varepsilon _{l}\left\langle 2\right\vert \vec{d%
}.\vec{e}_{l}\left\vert 1\right\rangle \hat{a}_{l}^{+}\exp \left( -i\vec{k}.%
\hat{\vec{R}}\right) \left\vert 2\right\rangle _{ii}\left\langle
1\right\vert +h.c.
\end{equation*}

\subsection{Master equation}

The master equation is developed in the reduced density matrix approach $%
\sigma =Tr_{V}(\rho )$. The evolution of the density $\rho $ is%
\begin{equation}
i\hbar \frac{d\rho }{dt}=\left[ H,\rho \right]
\end{equation}%
which leads to write%
\begin{equation}
i\hbar \frac{d\sigma }{dt}=Tr_{V}\left[ H,\rho \right]
\end{equation}%
We notice that $Tr_{V}\left[ H_{V},\rho \right] =0$ and $Tr_{V}\left[
H_{Ai},\rho \right] =\left[ H_{Ai},\sigma \right] $, which leads%
\begin{equation}
i\hbar \frac{d\sigma }{dt}=\left[ H_{A},\sigma \right] +Tr_{R}\left[
H_{AV},\rho \right]
\end{equation}

In the frame of the interaction representation%
\begin{equation}
\widetilde{C}=\exp \left( i\left( H_{A}+H_{V}\right) t/\hslash \right) C\exp
\left( -i\left( H_{A}+H_{V}\right) t/\hslash \right)
\end{equation}%
the evolution of the density matrix $\widetilde{\rho }$ is now%
\begin{equation}
i\hbar \frac{d\widetilde{\rho }(t)}{dt}=\left[ \widetilde{H}_{AV}(t),%
\widetilde{\rho }(t)\right]
\end{equation}%
which can be writen%
\begin{equation}
i\hbar \left( \widetilde{\rho }(t)-\widetilde{\rho }(0)\right) =\int_{0}^{t}%
\left[ \widetilde{H}_{AV}(t^{\prime }),\widetilde{\rho }(t^{\prime })\right]
dt^{\prime }
\end{equation}%
The evolution of the reduced density matrix $\widetilde{\sigma }$ is%
\begin{equation}
i\hbar \frac{d\widetilde{\sigma }(t)}{dt}=Tr_{V}\left[ \widetilde{H}_{AV}(t),%
\widetilde{\rho }(t)\right]
\end{equation}%
which can also be writen as%
\begin{equation}
\frac{d\widetilde{\sigma }(t)}{dt}=-\frac{1}{\hslash ^{2}}\left\{ Tr_{V}%
\left[ \widetilde{H}_{AV}(t) ,\widetilde{\rho }\left( 0\right) %
\right] +Tr_{V}\int_{0}^{t}\left[ \widetilde{H}_{AV}(t),\left[ \widetilde{H}%
_{AV}(t^{\prime }),\widetilde{\rho }(t^{\prime })\right] \right] dt^{\prime
}\right\}
\end{equation}%
By considering the change of variable $t^{\prime }=t-\tau $, with $\tau $\
varying from 0 to $t$%
\begin{equation}
\frac{d\widetilde{\sigma }(t)}{dt}=-\frac{1}{\hslash ^{2}}\left\{ Tr_{V}%
\left[ \widetilde{H}_{AV}(t) ,\widetilde{\rho }\left( 0\right) %
\right] +Tr_{R}\int_{0}^{t}\left[ \widetilde{H}_{AV}(t),\left[ \widetilde{H}%
_{AV}(t-\tau ),\widetilde{\rho }(t-\tau )\right] \right] d\tau \right\}
\end{equation}

\subsection{Born-Markov approximation}

We assume that at $t=0$ the electromagnetic field is the vacuum radiation $%
\left\vert 0\right\rangle _{VV}\left\langle 0\right\vert $.\ More we assume
the density matrix at the initial time, $\widetilde{\rho }(0)$, can be
factorizable, meaning there is no correlation between the atoms and the
vacuum field%
\begin{equation}
\widetilde{\rho }(0)=\widetilde{\sigma }(0)\otimes \left\vert 0\right\rangle
_{VV}\left\langle 0\right\vert
\end{equation}

To go further we make the Born-Markov approximation

\begin{itemize}
\item The Born approximation consists to consider that the atoms interact
always with the vacuum field an that no correlation appears between the
atoms and the vacuum field,meaning%
\begin{equation}
\widetilde{\rho }(t)=\widetilde{\sigma }(t)\otimes \left\vert 0\right\rangle
_{VV}\left\langle 0\right\vert
\end{equation}

\item The Markov approximation (or short memory) consists to replace $%
\widetilde{\sigma }(t-\tau )$\ par $\widetilde{\sigma }(t)$.\ We have%
\begin{equation}
\frac{d\widetilde{\sigma }(t)}{dt}=-\frac{1}{\hslash ^{2}}Tr_{V}\int_{0}^{t}%
\left[ \widetilde{H}_{AV}(t),\left[ \widetilde{H}_{AV}(t-\tau ),\widetilde{%
\sigma }(t-\tau )\otimes \left\vert 0\right\rangle _{VV}\left\langle
0\right\vert \right] \right] d\tau
\end{equation}%
\begin{equation}
\frac{d\widetilde{\sigma }(t)}{dt}=-\frac{1}{\hslash ^{2}}Tr_{V}\int_{0}^{t}%
\left[ \widetilde{H}_{AV}(t),\left[ \widetilde{H}_{AV}(t-\tau ),\widetilde{%
\sigma }(t)\otimes \left\vert 0\right\rangle _{VV}\left\langle 0\right\vert %
\right] \right] d\tau
\end{equation}
\end{itemize}

\subsection{Two-atom master equation}

By considering the interaction hamiltonian in the rotating wave
approximation, the master equation becomes%
\begin{eqnarray}
\frac{d\widetilde{\sigma }(t)}{dt} &=&-\frac{1}{\hbar ^{2}}\overset{}{%
\underset{i,j}{\sum }}\sum\limits_{l}\frac{\hbar ck}{2\varepsilon _{0}L^{3}}%
\left\vert \left\langle 2\right\vert \vec{d}.\vec{e}_{l}\left\vert
1\right\rangle \right\vert ^{2} \\
&&\int_{0}^{t}[\left\vert 2,i\right\rangle \left\langle 1,i\right\vert
\otimes \left\vert 1,j\right\rangle \left\langle 2,j\right\vert \widetilde{%
\sigma }(t)\exp \left( i\vec{k}.\left( \vec{R}_{i}-%
\vec{R}_{j}\right) \right) \exp \left( i\left( \omega
_{0}-ck\right) \tau \right) +  \notag \\
&&+\widetilde{\sigma }(t)\left\vert 2,i\right\rangle \left\langle
1,i\right\vert \otimes \left\vert 1,j\right\rangle \left\langle
2,j\right\vert \exp \left( i\vec{k}.\left( \vec{R}_{i}-%
\vec{R}_{j}\right) \right) \exp \left( -i\left( \omega
_{0}-ck\right) \tau \right)  \notag \\
&&-2\left\vert 1,i\right\rangle \left\langle 2,i\right\vert \widetilde{%
\sigma }(t)\left\vert 2,j\right\rangle \left\langle 1,j\right\vert \exp
\left( -i\vec{k}.\left( \vec{R}_{i}-\vec{R}_{j}\right)
\right) \cos \left( \left( \omega _{0}-ck\right) \tau \right) ]d\tau  \notag
\end{eqnarray}

with%
\begin{equation}
\sum\limits_{l}\left\vert \vec{d}.\vec{e}_{l}\right\vert ^{2}\equiv \left( 
\frac{L}{2\pi }\right) ^{3}\int k^{2}dkd\Omega _{\vec{k}}\left( \left\vert 
\vec{d}.\vec{e}_{l_{1}}\right\vert ^{2}+\left\vert \vec{d}.\vec{e}%
_{l_{2}}\right\vert ^{2}\right)
\end{equation}%
where $\vec{e}_{l_{1}}$\ et $\vec{e}_{l_{2}}$\ are two orthogonal
polarisations for the mode $\vec{k}$ : $\left\vert \vec{d}.\vec{e}%
_{l_{1}}\right\vert ^{2}+\left\vert \vec{d}.\vec{e}_{l_{2}}\right\vert
^{2}=\left\vert \vec{d}\right\vert ^{2}-\left\vert \frac{\vec{d}.%
\vec{k}}{k}\right\vert ^{2}$. The integration of the angular part 
$\int \left[ \left\vert \vec{d}\right\vert ^{2}-\left\vert \frac{\vec{d}.%
\vec{k}}{k}\right\vert ^{2}\right] d\Omega _{\vec{k}}=$ $2\pi
\left\vert \vec{d}\right\vert ^{2}\left( 2-\int_{0}^{\pi }\cos ^{2}\theta
\sin \theta d\theta \right) =\frac{8\pi }{3}\left\vert \vec{d}\right\vert
^{2}$\ gives%
\begin{equation}
\sum\limits_{l}\left\vert \vec{d}.\vec{e}_{l}\right\vert ^{2}\equiv \left( 
\frac{L}{2\pi }\right) ^{3}\frac{8\pi }{3}\left\vert \vec{d}\right\vert
^{2}\int k^{2}dk
\end{equation}%
By introducing the operators%
\begin{eqnarray*}
r_{i}^{+} &=&\left\vert 2,i\right\rangle \left\langle 1,i\right\vert \\
r_{i}^{-} &=&\left\vert 1,i\right\rangle \left\langle 2,i\right\vert
\end{eqnarray*}%
we have the master equation%
\begin{eqnarray}
\frac{d\widetilde{\sigma }(t)}{dt} &=&-\frac{1}{\hbar ^{2}}\frac{1}{8\pi ^{3}%
}\sum\limits_{l}\underset{i,j}{\sum }\int \frac{\hbar ck}{2\varepsilon _{0}}%
\left\vert \left\langle 2\right\vert \vec{d}.\vec{e}_{l}\left\vert
1\right\rangle \right\vert ^{2} \\
&&\int_{0}^{t}\left[ [r_{i}^{+}r_{j}^{-},\widetilde{\sigma }(t)]_{+}\exp
\left( i\vec{k}.\left( \vec{R}_{i}-\vec{R}_{j}\right)
\right) \cos \left( \left( \omega _{0}-ck\right) \tau \right) \right.  \notag
\\
&&-2r_{j}^{-}\widetilde{\sigma }(t)r_{i}^{+}\exp \left( i\vec{k}.\left( 
\vec{R}_{i}-\vec{R}_{j}\right) \right) \cos \left(
\left( \omega _{0}-ck\right) \tau \right)  \notag \\
&&\left. +i[r_{i}^{+}r_{j}^{-},\widetilde{\sigma }(t)]\exp \left( i\vec{k}%
.\left( \vec{R}_{i}-\vec{R}_{j}\right) \right) \sin
\left( \left( \omega _{0}-ck\right) \tau \right) \right] d\tau d^{3}\vec{k} 
\notag
\end{eqnarray}

The terms, non considered in the above equation, corresponding to the non
resonant part of the hamiltonian are%
\begin{eqnarray}
\frac{d\widetilde{\sigma }^{(NR1)}(t)}{dt} &=&-\frac{1}{\hbar ^{2}}\frac{1}{%
8\pi ^{3}}\sum\limits_{l}\underset{i,j}{\sum }\int \frac{\hbar ck}{%
2\varepsilon _{0}}\left\vert \left\langle 2\right\vert \vec{d}.\vec{e}%
_{l}\left\vert 1\right\rangle \right\vert ^{2} \\
&&\int_{0}^{t}\left[ [r_{i}^{-}r_{j}^{+},\widetilde{\sigma }(t)]_{+}\exp
\left( i\vec{k}.\left( \vec{R}_{i}-\vec{R}_{j}\right)
\right) \cos \left( \left( -\omega _{0}-ck\right) \tau \right) \right. 
\notag \\
&&-2r_{j}^{+}\widetilde{\sigma }(t)r_{i}^{-}\exp \left( i\vec{k}.\left( 
\vec{R}_{i}-\vec{R}_{j}\right) \right) \cos \left(
\left( -\omega _{0}-ck\right) \tau \right)  \notag \\
&&\left. +i[r_{i}^{-}r_{j}^{+},\widetilde{\sigma }(t)]\exp \left( i\vec{k}%
.\left( \vec{R}_{i}-\vec{R}_{j}\right) \right) \sin
\left( \left( -\omega _{0}-ck\right) \tau \right) \right] d\tau d^{3}\vec{k}
\notag
\end{eqnarray}

There is also  crossed terms of the resonant and non resonant parts
of the hamiltonian containing $\exp \left( \pm 2i\omega
_{0}t\right) $ phase evolution and that we are going to neglect because of their off-resonant nature. 
 The conditions of validity for this approximation
correspond to a slow variation of the density matrix, $\sigma $, compared to
the period, $2\pi /\omega _{0}$.

We choose the axis, $z$, as $\vec{R}_{i}-\vec{R}_{j}$, 
$\left( \vec{e}_{z}=\left( \vec{R}_{i}-\vec{%
R}_{j}\right) /\left\vert \vec{R}_{i}-\vec{R}%
_{j}\right\vert \right) $. The emission mode, $\vec{k}$, is
characterized by the angles $\theta ^{\prime }$ and $\varphi ^{\prime }$,
and the polarisations $\vec{e}_{l1}=\sin \varphi ^{\prime }\vec{e}_{x}-\cos
\varphi ^{\prime }\vec{e}_{y}$ and $\vec{e}_{l2}=-\cos \theta ^{\prime }\cos
\varphi ^{\prime }\vec{e}_{x}-\cos \theta ^{\prime }\sin \varphi ^{\prime }%
\vec{e}_{y}+\sin \theta ^{\prime }\vec{e}_{z}$. The spherical coordonates of
the transition dipole vector $\vec{d}$ are $\left\{ d,\theta _{ij}=\theta
,\varphi _{ij}=\varphi \right\} $. We have%
\begin{equation*}
\vec{k}.\left( \vec{R}_{i}-\vec{R}_{j}\right)
=kR_{ij}\cos \theta ^{\prime }
\end{equation*}%
\begin{eqnarray*}
\sum\limits_{l}\left\vert \left\langle 2\right\vert \vec{d}.\vec{e}%
_{l}\left\vert 1\right\rangle \right\vert ^{2} &=&\left\vert \left\langle
2\right\vert \vec{d}\left\vert 1\right\rangle \right\vert ^{2}-\left\vert
\left\langle 2\right\vert \frac{\vec{d}.\vec{k}}{k}\left\vert
1\right\rangle \right\vert ^{2} \\
\left\vert \left\langle 2\right\vert \frac{\vec{d}.\vec{k}}{k}%
\left\vert 1\right\rangle \right\vert ^{2} &=&\left\vert \left\langle
2\right\vert \vec{d}\left\vert 1\right\rangle \right\vert ^{2}\left( \cos
\theta \cos \theta ^{\prime }+\sin \theta \sin \theta ^{\prime }\cos \left(
\varphi -\varphi ^{\prime }\right) \right) ^{2}
\end{eqnarray*}%
For large $t$, we can write%
\begin{equation}
\int_{0}^{t\approx \infty }\left[ {\exp }i\left( ck-\omega _{0}\right) \tau %
\right] d\tau =\frac{\pi }{c}\delta \left( k-\frac{\omega _{0}}{c}\right) +%
\frac{i}{c}PP\frac{1}{k-\frac{\omega _{0}}{c}}
\end{equation}%
\begin{eqnarray}
\frac{d\widetilde{\sigma }(t)}{dt} &=&-\frac{1}{\hbar ^{2}}\frac{1}{8\pi ^{3}%
}\underset{i,j}{\sum }\int \frac{\hbar ck}{2\varepsilon _{0}}\left\vert
\left\langle 2\right\vert \vec{d}\left\vert 1\right\rangle
\right\vert ^{2} \\
&&\left( 1-\left( \cos \theta \cos \theta ^{\prime }+\sin \theta \sin \theta
^{\prime }\cos \left( \varphi -\varphi ^{\prime }\right) \right) ^{2}\right)
\notag \\
&&\left[ \left[ [r_{i}^{+}r_{j}^{-},\widetilde{\sigma }(t)]_{+}-2r_{j}^{-}%
\widetilde{\sigma }(t)r_{i}^{+}\right] \right. \exp \left( ikR_{ij}\cos
\theta ^{\prime }\right) \frac{\pi }{c}\delta \left( k-k_{0}\right)  \notag
\\
&&\left. +i[r_{i}^{+}r_{j}^{-},\widetilde{\sigma }(t)]\exp \left(
ikR_{ij}\cos \theta ^{\prime }\right) \frac{1}{c}PP\frac{1}{k-k_{0}}\right]
k^{2}dk\sin \theta ^{\prime }d\theta ^{\prime }d\varphi ^{\prime }  \notag
\end{eqnarray}%
\begin{eqnarray}
\frac{d\widetilde{\sigma }(t)}{dt} &=&-\frac{1}{\hbar ^{2}}\frac{1}{8\pi ^{3}%
}\underset{i,j}{\sum }\int_{k=0}^{+\infty }\int_{u=-1}^{+1}\frac{\hbar ck}{%
2\varepsilon _{0}}\left\vert \left\langle 2\right\vert \vec{d}%
\left\vert 1\right\rangle \right\vert ^{2}\pi \\
&&\left( 1+\cos ^{2}\theta +\left( 1-3\cos ^{2}\theta \right) u^{2}\right) 
\notag \\
&&\left[ \left[ [r_{i}^{+}r_{j}^{-},\widetilde{\sigma }(t)]_{+}-2r_{j}^{-}%
\widetilde{\sigma }(t)r_{i}^{+}\right] \right. \exp \left( ikR_{ij}u\right) 
\frac{\pi }{c}\delta \left( k-k_{0}\right)  \notag \\
&&\left. +i[r_{i}^{+}r_{j}^{-},\widetilde{\sigma }(t)]\exp \left(
ikR_{ij}u\right) \frac{1}{c}PP\frac{1}{k-k_{0}}\right] k^{2}dkdu  \notag
\end{eqnarray}

We should add the contribution of the non-resonant terms%
\begin{eqnarray}
\frac{d\widetilde{\sigma }^{(NR1)}(t)}{dt} &=&-\frac{1}{\hbar ^{2}}\frac{1}{%
8\pi ^{3}}\underset{i,j}{\sum }\int_{k=0}^{+\infty }\int_{u=-1}^{+1}\frac{%
\hbar ck}{2\varepsilon _{0}}\left\vert \left\langle 2\right\vert 
\vec{d}\left\vert 1\right\rangle \right\vert ^{2}\pi \\
&&\left( 1+\cos ^{2}\theta +\left( 1-3\cos ^{2}\theta \right) u^{2}\right) 
\notag \\
&&\left[ [r_{i}^{-}r_{j}^{+},\widetilde{\sigma }(t)]_{+}\right. \exp \left(
ikR_{ij}u\right) \frac{\pi }{c}\delta \left( k+k_{0}\right)  \notag \\
&&\left. +i[r_{i}^{-}r_{j}^{+},\widetilde{\sigma }(t)]\exp \left(
ikR_{ij}u\right) \frac{1}{c}PP\frac{1}{k+k_{0}}\right] k^{2}dkdu  \notag
\end{eqnarray}%

By changing $k$ in $-k$ in the integral, we have%
\begin{eqnarray}
\frac{d\widetilde{\sigma }^{(NR1)}(t)}{dt} &=&-\frac{1}{\hbar ^{2}}\frac{1}{%
8\pi ^{3}}\underset{i,j}{\sum }\int_{k=-\infty }^{0}\int_{u=-1}^{+1}\frac{%
\hbar ck}{2\varepsilon _{0}}\left\vert \left\langle 2\right\vert 
\vec{d}\left\vert 1\right\rangle \right\vert ^{2}\pi  \\
&&\left( 1+\cos ^{2}\theta +\left( 1-3\cos ^{2}\theta \right) u^{2}\right)  
\notag \\
&&\left[ [r_{i}^{-}r_{j}^{+},\widetilde{\sigma }(t)]_{+}\right. \exp \left(
-ikR_{ij}u\right) \frac{\pi }{c}\delta \left( -k+k_{0}\right)   \notag \\
&&\left. +i[r_{i}^{-}r_{j}^{+},\widetilde{\sigma }(t)]\exp \left(
-ikR_{ij}u\right) \frac{1}{c}PP\frac{1}{k-k_{0}}\right] k^{2}dkdu  \notag
\end{eqnarray}%
The operators $r_{i}^{\pm }$ and $r_{j}^{\pm }$ commute when $i\neq j$. By
inverting the indices $i$ and $j$, we have to replace $u$ by $-u$%
\begin{eqnarray}
\frac{d\widetilde{\sigma }^{(NR1)}(t)}{dt} &=&-\frac{1}{\hbar ^{2}}\frac{1}{%
8\pi ^{3}}\underset{i,j}{\sum }\int_{k=-\infty }^{0}\int_{u=-1}^{+1}\frac{%
\hbar ck}{2\varepsilon _{0}}\left\vert \left\langle 2\right\vert 
\vec{d}\left\vert 1\right\rangle \right\vert ^{2}\pi  \\
&&\left( 1+\cos ^{2}\theta +\left( 1-3\cos ^{2}\theta \right) u^{2}\right)  
\notag \\
&&\left[ [r_{i}^{+}r_{j}^{-},\widetilde{\sigma }(t)]_{+}\right. \exp \left(
ikR_{ij}u\right) \frac{\pi }{c}\delta \left( k-k_{0}\right)   \notag \\
&&\left. +i[r_{i}^{+}r_{j}^{-},\widetilde{\sigma }(t)]\exp \left(
ikR_{ij}u\right) \frac{1}{c}PP\frac{1}{k-k_{0}}\right] k^{2}dkdu  \notag
\end{eqnarray}%
By taking into account both resonant and non-resonant terms, we have%
\begin{eqnarray}
\frac{d\widetilde{\sigma }(t)}{dt} &=&-\Gamma \underset{i}{\sum }\left[
[r_{i}^{+}r_{i}^{-},\widetilde{\sigma }(t)]_{+}-2r_{i}^{-}\widetilde{\sigma }%
(t)r_{i}^{+}\right]  \\
&&-\frac{1}{\hbar ^{2}}\frac{1}{8\pi ^{3}}\underset{i\neq ,j}{\sum }%
\int_{k=-\infty }^{+\infty }\int_{u=-1}^{+1}\frac{\hbar ck}{2\varepsilon _{0}%
}\left\vert \left\langle 2\right\vert \vec{d}\left\vert
1\right\rangle \right\vert ^{2}\pi   \notag \\
&&\left( 1+\cos ^{2}\theta +\left( 1-3\cos ^{2}\theta \right) u^{2}\right)  
\notag \\
&&\left[ [r_{i}^{+}r_{j}^{-},\widetilde{\sigma }(t)]_{+}\right. \exp \left(
ikR_{ij}u\right) \frac{\pi }{c}\delta \left( k-k_{0}\right)   \notag \\
&&\left. +i[r_{i}^{+}r_{j}^{-},\widetilde{\sigma }(t)]\exp \left(
ikR_{ij}u\right) \frac{1}{c}PP\frac{1}{k-k_{0}}\right] k^{2}dkdu  \notag
\end{eqnarray}%
We do not consider the term of the principal part $PP$ in the integral for $%
i=j$.\ This term is already taken into account in the energy of the
different levels in the Lamb shift calculation. We use the equation%
\begin{equation*}
\int_{-\infty }^{+\infty }\exp \left( i\left( k-k_{0}\right) R_{ij}u\right) 
\left[ \pi \delta \left( k-k_{0}\right) +iPP\frac{1}{k-k_{0}}\right] dk=2\pi
\Theta \left( R_{ij}u\right) 
\end{equation*}%
with $\Theta \left( x\right) =1$ is the Heaviside function with for $x>0$ , $%
\Theta \left( x\right) =1$, for $x=0$, $\Theta \left( x\right) =1/2$, for $%
x<0$, $\Theta \left( x\right) =0$. For integrating, we notice also that $%
k^{3}\exp \left( ikR_{ij}u\right) =\frac{i}{R_{ij}^{3}}\frac{\partial ^{3}}{%
\partial u^{3}}\exp \left( ikR_{ij}u\right) $. We can write%
\begin{eqnarray}
\frac{d\widetilde{\sigma }(t)}{dt} &=&-\Gamma \underset{i}{\sum }\left[
[r_{i}^{+}r_{i}^{-},\widetilde{\sigma }(t)]_{+}-2r_{i}^{-}\widetilde{\sigma }%
(t)r_{i}^{+}\right]  \\
&&-\frac{1}{\hbar }\frac{1}{4\pi \varepsilon _{0}}\underset{i\neq j}{\sum }%
\left\vert \left\langle 2\right\vert \vec{d}\left\vert
1\right\rangle \right\vert ^{2}  \notag \\
&&\left\{ \left( 1-\cos ^{2}\theta \right) \frac{\left( -i\right) k_{0}^{2}}{%
R_{ij}}\left[ r_{i}^{+}r_{j}^{-}\widetilde{\sigma }(t)\exp \left(
ik_{0}R_{ij}\right) \right. \right.   \notag \\
&&\left. -\widetilde{\sigma }(t)r_{i}^{+}r_{j}^{-}\exp \left(
-ik_{0}R_{ij}\right) \right]   \notag \\
&&-\left( 1-3\cos ^{2}\theta \right) \frac{-k_{0}}{R_{ij}^{2}}\left[
r_{i}^{+}r_{j}^{-}\widetilde{\sigma }(t)\exp \left( ik_{0}R_{ij}\right)
\right.   \notag \\
&&\left. +\widetilde{\sigma }(t)r_{i}^{+}r_{j}^{-}\exp \left(
-ik_{0}R_{ij}\right) \right]   \notag \\
&&+\left( 1-3\cos ^{2}\theta \right) \frac{i}{R_{ij}^{3}}\left[
r_{i}^{+}r_{j}^{-}\widetilde{\sigma }(t)\exp \left( ik_{0}R_{ij}\right)
\right.   \notag \\
&&\left. \left. -\widetilde{\sigma }(t)r_{i}^{+}r_{j}^{-}\exp \left(
-ik_{0}R_{ij}\right) \right] \right\}   \notag
\end{eqnarray}

Finally,
the equation can be written%
\begin{eqnarray}
\frac{d\widetilde{\sigma }(t)}{dt} &=&-\Gamma \underset{i}{\sum }\left[
[r_{i}^{+}r_{i}^{-},\widetilde{\sigma }(t)]_{+}-2r_{i}^{-}\widetilde{\sigma }%
(t)r_{i}^{+}\right]  \notag \\
&&-\frac{1}{\hbar }\frac{1}{4\pi \varepsilon _{0}}\underset{i\neq j}{\sum }%
\left\vert \left\langle 2\right\vert \vec{d}\left\vert
1\right\rangle \right\vert ^{2}  \notag \\
&&\left\{ \left\{ \left( 1-\cos ^{2}\theta \right) \frac{k_{0}^{2}}{R_{ij}}%
\sin \left( k_{0}R_{ij}\right) \right. \right.  \notag \\
&&+\left( 1-3\cos ^{2}\theta \right) \frac{k_{0}}{R_{ij}^{2}}\cos \left(
k_{0}R_{ij}\right)  \notag \\
&&\left. -\left( 1-3\cos ^{2}\theta \right) \frac{1}{R_{ij}^{3}}\sin \left(
k_{0}R_{ij}\right) \right\}  \notag \\
&&\left[ r_{i}^{+}r_{j}^{-},\widetilde{\sigma }(t)\right] _{+}  \notag \\
&&-i\left\{ \left( 1-\cos ^{2}\theta \right) \frac{k_{0}^{2}}{R_{ij}}\cos
\left( k_{0}R_{ij}\right) \right.  \notag \\
&&-\left( 1-3\cos ^{2}\theta \right) \frac{k_{0}}{R_{ij}^{2}}\sin \left(
k_{0}R_{ij}\right)  \notag \\
&&\left. -\left( 1-3\cos ^{2}\theta \right) \frac{1}{R_{ij}^{3}}\cos \left(
k_{0}R_{ij}\right) \right\}  \notag \\
&&\left. \left[ r_{i}^{+}r_{j}^{-},\widetilde{\sigma }(t)\right] \right\}
\end{eqnarray}%

\subsection{\protect\bigskip Dipole-dipole interaction}

The master equation  can be written%
\begin{equation*}
i\hbar \frac{d\widetilde{\sigma }(t)}{dt}= -\Gamma \underset{i}{\sum }\left[
[r_{i}^{+}r_{i}^{-},\widetilde{\sigma }(t)]_{+}-2r_{i}^{-}\widetilde{\sigma }%
(t)r_{i}^{+}\right] +  \underset{i\neq j}{\sum } \left[
\widetilde{H}_{ij},\widetilde{\sigma }(t)\right] -\frac{i}{2}\left( 
\widetilde{\Gamma }_{ij}\widetilde{\sigma }(t)+\widetilde{\sigma }(t)%
\widetilde{\Gamma }_{ij}\right) 
\end{equation*}%
with%
\begin{eqnarray}
\widetilde{H}_{ij} &=&\frac{1}{4\pi \varepsilon _{0}} \left\vert \left\langle 2\right\vert \vec{d}\left\vert
1\right\rangle \right\vert ^{2}r_{i}^{+}r_{j}^{-}  \notag \\
&&\left\{ -\left( 1-\cos ^{2}\theta \right) \frac{k_{0}^{2}}{R_{ij}}\cos
\left( k_{0}R_{ij}\right) \right.   \notag \\
&&+\left( 1-3\cos ^{2}\theta \right) \frac{k_{0}}{R_{ij}^{2}}\sin \left(
k_{0}R_{ij}\right)   \notag \\
&&\left. +\left( 1-3\cos ^{2}\theta \right) \frac{1}{R_{ij}^{3}}\cos \left(
k_{0}R_{ij}\right) \right\} 
\label{eq_ham_finale}
\end{eqnarray}%

and%
\begin{eqnarray}
\frac{\widetilde{\Gamma }_{ij}}{2}= &&-i\frac{1}{4\pi \varepsilon _{0}}%
\left\vert \left\langle 2\right\vert 
\vec{d}\left\vert 1\right\rangle \right\vert
^{2}r_{i}^{+}r_{j}^{-}  \notag \\
&&\Big\{ \left( 1-\cos ^{2}\theta \right) \frac{k_{0}^{2}}{R_{ij}}%
\sin \left( k_{0}R_{ij}\right)   \notag \\
&&+\left( 1-3\cos ^{2}\theta \right) \frac{k_{0}}{R_{ij}^{2}}\cos \left(
k_{0}R_{ij}\right)   \notag \\
&& -\left( 1-3\cos ^{2}\theta \right) \frac{1}{R_{ij}^{3}}\sin \left(
k_{0}R_{ij}\right) \Big\}   \notag
\end{eqnarray}

$\widetilde{H}_{ij}$ can be interpreted in terms of interaction between two
oscillating electric dipoles, $\vec{\mu }_{i}$ and $%
\vec{\mu }_{j}$, at a distance $R_{ij}\vec{n}_{ij}$.
This dipole-dipole interaction between two
atoms can be interpreted as a collective Lamb shift due to the
dipole-dipole interaction. The term in $R^{-3}$ correspond to the classical
dipole-dipole interaction%
The term in $1/R$ correspond to the delayed terms of the interaction.

\ The $\widetilde{\Gamma }_{ij}$
terms contains the spontaneous emission by individual
atoms.

\end{document}